\documentclass[useAMS,usenatbib,usegraphicx]{mn2e}

\usepackage{amssymb}

\newcommand{\var}{\mathop{\rm Var}\nolimits}

\newcommand{\trace}{\mathop{\rm Tr}\nolimits}
\newcommand{\expect}{\mathbb{E}}
\newcommand{\FAP}{{\rm FAP}}

\title[Planets orbiting GJ581]{The impact of red noise in radial velocity planet searches:
Only three planets orbiting GJ581?}

\author[R.V.~Baluev]{Roman V. Baluev\thanks{E-mail: roman@astro.spbu.ru}\\
Central (Pulkovo) Astronomical Observatory of Russian Academy of Sciences, Pulkovskoje sh. 65/1, St Petersburg 196140, Russia\\
Sobolev Astronomical Institute, St Petersburg State University, Universitetskij prospekt
28, Petrodvorets, St Petersburg 198504, Russia}

\begin{document}

\date{
      Received 2012 November 21;
      in original form 2012 September 14}

\pagerange{\pageref{firstpage}--\pageref{lastpage}} \pubyear{2012}

\maketitle

\label{firstpage}

\begin{abstract}
We perform a detailed analysis of the latest HARPS and Keck radial velocity data for the
planet-hosting red dwarf GJ581, which attracted a lot of attention in recent time. We show
that these data contain important correlated noise component (``red noise'') with the
correlation timescale of the order of $10$~days. This red noise imposes a lot of
misleading effects while we work in the traditional white-noise model. To eliminate these
misleading effects, we propose a maximum-likelihood algorithm equipped by an extended
model of the noise structure. We treat the red noise as a Gaussian random process with
exponentially decaying correlation function.

Using this method we prove that: (i) planets \emph{b} and \emph{c} do exist in this
system, since they can be independently detected in the HARPS and Keck data, and
regardless of the assumed noise models; (ii) planet \emph{e} can also be confirmed
independently by the both datasets, although to reveal it in the Keck data it is mandatory
to take the red noise into account; (iii) the recently announced putative planets \emph{f}
and \emph{g} are likely just illusions of the red noise; (iv) the reality of the planet
candidate GJ581~\emph{d} is questionable, because it cannot be detected from the Keck
data, and its statistical significance in the HARPS data (as well as in the combined
dataset) drops to a marginal level of $\sim 2\sigma$, when the red noise is taken into
account.

Therefore, the current data for GJ581 really support existence of no more than four (or
maybe even only three) orbiting exoplanets. The planet candidate GJ581~\emph{d} requests
serious observational verification.
\end{abstract}


\section{Introduction}
The multi-planet extrasolar system hosted by the red dwarf GJ581 has attracted a lot of
interest in the past few years. The concise history of planet detections for this system
is as follows. The first planet \emph{b}, having orbital period of $5.37$~d and minimum
mass of $\sim 15 M_\oplus$, was reported by \citet{Bonfils05}. The two subsequent
super-Earths \emph{c} (with the orbital period of $12.9$~d and the minimum mass of $\sim 5
M_\oplus$) and \emph{d} (with the originally reported orbital period of $82$~d later
corrected to $67$~d and current minimum mass estimate of $6 M_\oplus$) were discovered by
\citet{Udry07}. Further, \citet{Mayor09} reported the detection of the smallest exoplanet
known so far, GJ581~\emph{e}, orbiting the host star each $3.15$~d and having the minimum
mass of only appoximately $2 M_\oplus$. All these discoveries were done on the basis of
the radial velocity data obtained with the famous HARPS spectrograph.

Later, the Keck planet-search team got involved. \citet{Vogt10} performed an analysis of
the combined HARPS and Keck measurements and claimed the detection of two more planets in
the system, \emph{f} and \emph{g}, orbiting the host star each $433$~d and $36.6$~d (and
having minimum masses of $\sim 7$ and $\sim 3 M_\oplus$). The last planet \emph{g} is
remarkable because it appears to reside in the middle of the predicted habitable zone for
this star. However, the reality of these two planets represents a subject of serious
debates in the recent time. \citet{Gregory11} remained uncertain about the existence of
these planets, based on his very detailed Bayesian analysis of the joint \citep{Mayor09}
and \citep{Vogt10} datasets. \citet{dosSantos12} basically agreed with this conclusion,
finding from the same combined dataset that the detection confidence probabilties for
these two planets are $96\%$ for planet \emph{g} and $98\%$ for planet \emph{f}. These
values are too high to be just neglected, but they are simultaneously too low to claim a
robust detection. \citet{Forveille11} claim in a recent preprint that newer HARPS data do
not support the existence of any planets beyond the four-planet model. Finally, in a very
recent paper \citep{Vogt12} the authors assert, on the basis of the HARPS data from
\citep{Forveille11}, that with the false-alarm probability of $\sim 4\%$ an extra $32$-day
planet should exist in this system, beyond the four-planet model.

Summarizing these investigations, we must admit that the reality of the last detected
planets is rather controversial. This uncertainty probably comes from some mysterious
interference between the HARPS and Keck data. Indeed, it follows from e.g.
\citep{Gregory11} and \citep{dosSantos12} that Keck data alone do not allow to detect more
than only \emph{two} planets \emph{b} and \emph{c}: all other planets seem to fall beyond
the detection power of this time series. Newest HARPS data alone allow for the robust
detection of \emph{four} planets (from \emph{b} to \emph{e}) and do not really support the
existence of the planets \emph{f} and \emph{g}. However, some additional variations can be
still detected when the both datasets are joined, and it is rather uncomfortable just to
ignore them.

In this paper we present an attempt to find a solution of this mystery. Our main idea
comes from our previous work \citep{Baluev11}, where we analysed available radial velocity
(RV) data for another planet-hosting red dwarf GJ876, and found that these data contain
significant correlated noise component, also called as ``red noise''. Traditional
statistical methods assume that the measurement errors are statistically independent,
implying that their frequency power spectrum is flat (thus the noise is ``white''). As we
have shown in \citep{Baluev11}, both HARPS and Keck radial velocity measurements of GJ876
demonstrate non-white power spectra with a clearly visible excess at longer periods, and
this non-whiteness is statistically significant. We found that a similar picture is often
seen in the periodograms of the GJ581 data (see, e.g., a lot of periodograms plotted by
\citealt{dosSantos12}). All this motivated us to investigate how the red noise could
affect the derived orbital configuration of this system.

The structure of the paper is as follows. First, in the Section~\ref{sec_rednoise} we
discuss the common undesired effects that the red noise might impose, and demonstrate how
it reveal itself in the GJ581 RV data. In Section~\ref{sec_reduct} we present a
maximum-likelihood algorithm that can perform a reduction of the red noise, based on its
full modeling. In the Section~\ref{sec_analysis} we perform a detailed analysis of the
latest radial velocity data for GJ581 taken from \citep{Vogt10} and \citep{Forveille11}.
We show how in the particular case of GJ581 the red noise creates fake RV variations, as
well as hides the true ones. We also give two best fitting orbital solutions for this
system, that take the red noise into account. In Section~\ref{sec_planet56}, we discuss
the reality of the putative fifth and sixth planets, based on our RV data analysis. In the
conclusive section of the paper, we discuss what global consequences the red noise implies
for the past exoplanetary data-analysis works and what it requests from the future ones.

\section{Red noise as a misleading agent}
\label{sec_rednoise}
The routinely used methods of astrostatistics are designed to deal with the data
containing uncorrelated noise. Such noise is also called white, because its frequency
spectrum is flat: its periodograms demonstrate approximately the same mean level when
averaged over different frequency segments of the same length.

In practice, however, the white-noise approximation may be poor. In particular, the noise
in photometric observations of exoplanetary transits is routinely red \citep{Pont06}. In
the radial velocity planet searches, it is also known that the RV noise is not necessarily
white, because it may demonstrate smaller level when averaged over larger timescales
\citep[e.g.][]{OToole08}. However, for the RV case this issue basically appears as rather
dark stuff with no routinely working practical solution known so far.

\begin{figure}
\includegraphics[width=84mm]{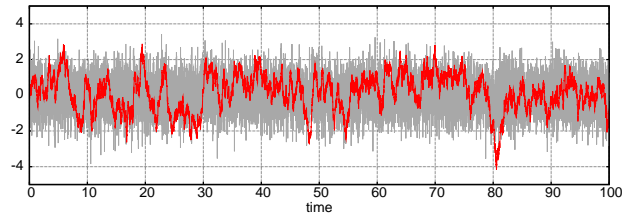}
\caption{Foreground curve shows a simulated example of the red Gaussian noise with correlation
function $e^{-|\tau|}$, while thick background band shows a simulated example of the white
Gaussian noise of the same variance.}
\label{fig_rednoise}
\end{figure}

Potential impact of the red noise on the results of the data analysis may be huge. The
correlated data usually carry smaller amount of information, as if their number was
smaller. Therefore, when our data contain correlated noise, various statistical
uncertainties are typically larger than we obtain based on the traditional white-noise
models. It is the first effect imposed by the red noise. Another, possibly even more
important effect, appears due to the non-uniform frequency spectrum of the red noise.
Basically, the red noise is able to generate fake periodicities that can be mistakenly
``detected'' by the white-noise algorithms. This is demonstrated in
Fig.~\ref{fig_rednoise}, where the simulated example of a correlated noise looks like a
bit noisy mixture of some illusive periodic signals. Moreover, these fake periodicities
may obscure real variations, keeping them undetected until some data-analysis tool that is
aware of the noise correlateness is applied.

\begin{figure}
\includegraphics[width=84mm]{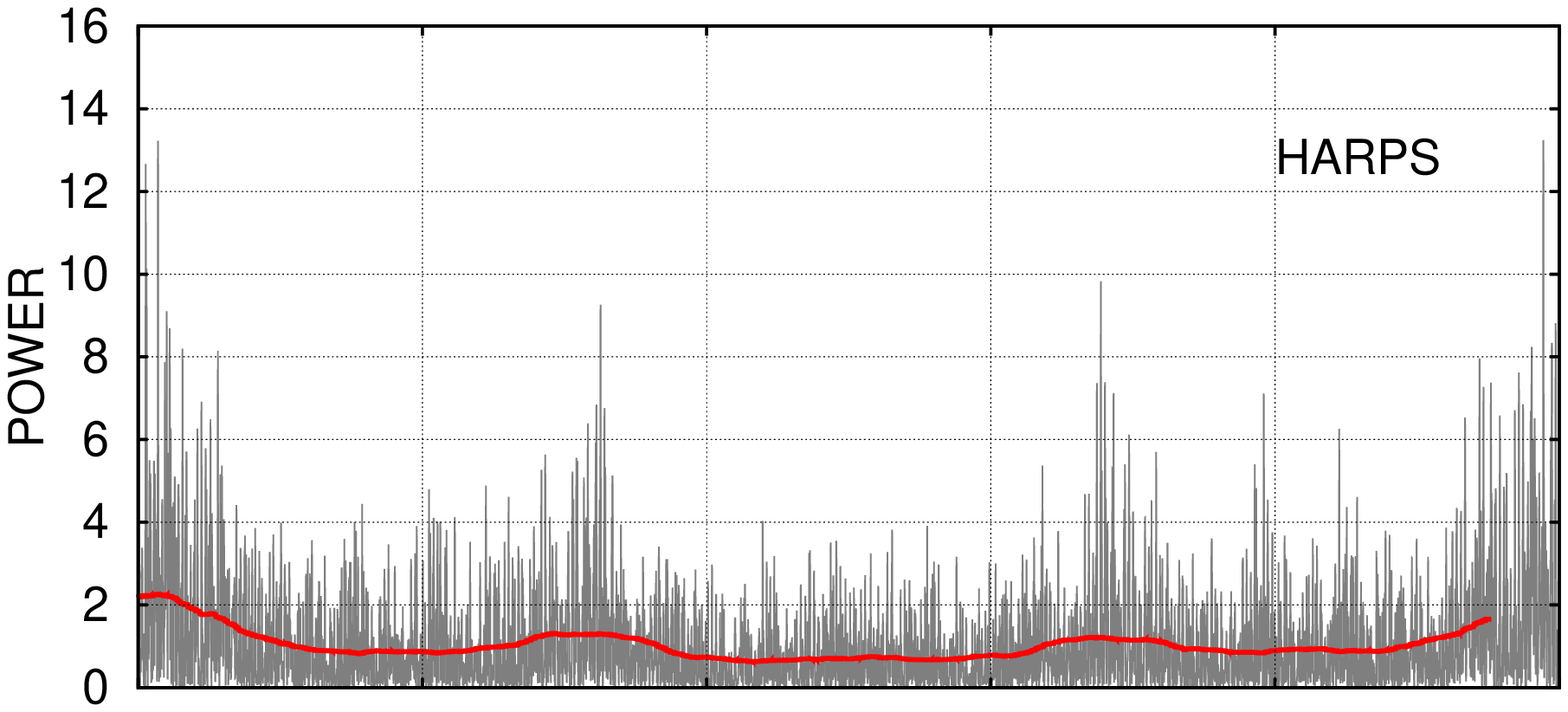}
\includegraphics[width=84mm]{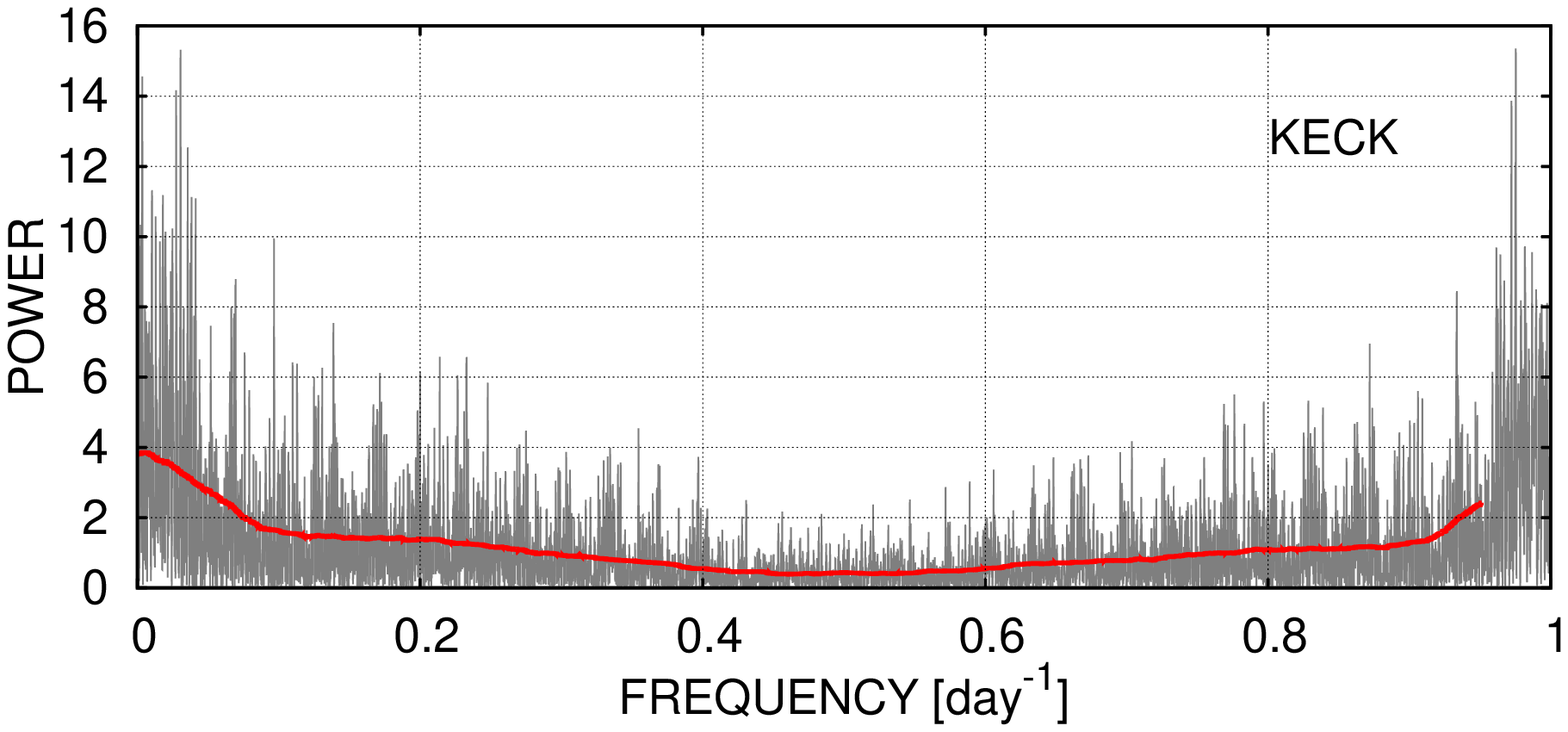}
\caption{The red noise in the HARPS and KECK RV data for GJ581 in the frequency domain.
We plot the residual periodograms that remain after elimination of the compound $4$-planet
signal. These periodograms are constructed separately for the HARPS and Keck data by means
of adding the probe sinusoidal signal to the RV model of only HARPS or only Keck dataset,
but still fitting such RV model jointly to the both datasets. Each value of these
periodograms represents the modified likelihood ratio statistic $\tilde Z$ defined in
\citep{Baluev08b}. The smoothed curves represent the moving average of the raw
periodograms obtained using the window of $0.1$~day$^{-1}$.}
\label{fig_rnGJ581}
\end{figure}

In the case of GJ581, the white-noise model of the RV data is definitely inadequate. As we
can see from Fig.~\ref{fig_rnGJ581}, the periodograms of the residual noise that remains
after elimination of the compound RV signal of 4 planets demonstrate clear excess of the
power at low frequencies ($\lesssim 0.1$~day$^{-1}$) a symmetric excess around $1$~day
period (emerging due to a strong diurnal aliasing), and a depression in the middle of the
segment. It is importaint that this power excess does not concentrate in any well-defined
discrete peaks; instead it is spread smoothly in a continuous frequency band. The both
periodograms for the HARPS and Keck data demonstrate a similar smoothed shape, although
the positions of individual high peaks have little common (meaning that all relevant
periodicities are not real). We may note that this picture looks very similar to the one
we have already seen for the GJ876 case \citep{Baluev11}.

\begin{figure}
\includegraphics[width=84mm]{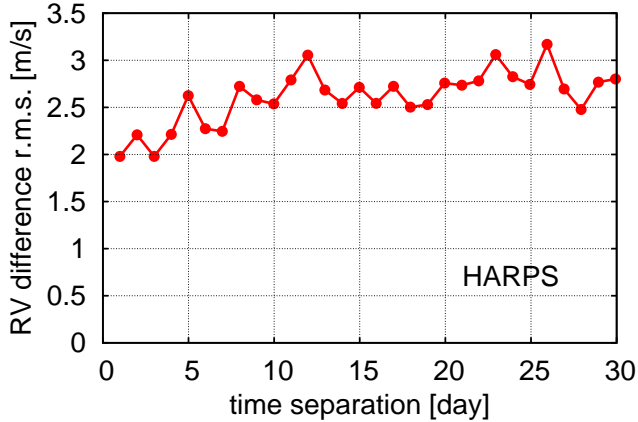}
\caption{The red noise in the HARPS RV data for GJ581 in the time domain. See text for the
detailed description.}
\label{fig_corrGJ581}
\end{figure}

The RV noise correlation can be also revealed in the time domain. This is easy to do for
the HARPS data thanks to the following favouring factors:
\begin{enumerate}
\item All HARPS measurements were done at almost the same sidereal time, implying that
the time separation between two arbitrary datapoints is usually very close to an integer
number of sidereal days.
\item There are a lot (up to $\sim 100$) of the HARPS RV pairs that have a small time
separation of only one or a few sidereal days.
\end{enumerate}

We run the following procedure. Given some integer $n$ from $1$ to $30$, we collect all
pairs of the HARPS measurements that are separated by $n$ sidereal days. For each such
pair we evaluate the difference of the 4-planet RV residuals corresponding to the
observations involved in this pair. Finally, for each group of the RV pairs with a given
time separation we evaluate the sample variance of this RV difference. How this variance
can help us to detect red noise? Assuming that the variance of the residual RV noise has
some constant value of $\sigma^2$ (which is not too far from the truth) and its
autocorrelation function is $R(\Delta t)$, the variance of the mentioned above RV
difference should be $2 \sigma^2 (1-R(\Delta t))$. Therefore, the graph of this variance
should basically represent a rescaled upside-down view of the noise correlation function.
We show this plot in Fig.~\ref{fig_corrGJ581}. We can see clear growing trend before the
time separation of $10$~days, and a saturation beyond this point. Basically, the HARPS RV
measurements have better relative precision at short timescales of up to $\sim 10$~days,
while at longer time separations they show larger random scatter.

Unfortunately, the distribution of the Keck data points is not regular enough, and
application of a similar procedure to the Keck dataset was not informative.

In this section we limit ourself by demonstration only, leaving the rigorous determination
of the red noise significance for further sections. However, it is already clear that we
may obtain trustable results concerning the GJ581 planetary system only if we utilize some
method of the data analysis that can properly deal with correlated noise.

\section{Maximum-likelihood reduction of the red noise}
\label{sec_reduct}
The method that we propose for the analysis of the data polluted by correlated noise
represents a generalization of the maximum-likelihood approach described in
\citep{Baluev08b}, which was already used in \citep{Baluev11}. The main idea of the method
is to construct some suitable correlational model of the noise in the RV data and then,
based on this model and on the Keplerian model of the RV curve, apply the
maximum-likelihood alorithm to estimate the parameters of the both models and the relevant
goodness of the fit.

Thus, first we should choose some realistic and simultaneously simple model of such
correlated noise. We assume that this noise is a Gaussian random process with some known
correlation function. This means that the full vector of our $N$ RV measurements $x_i$,
taken at the timings $t_i$, should follow a multivariate Gaussian distribution. The mean
of this distribution equal to the RV curve model $\mu(t_i,\btheta)$, and the relevant
variance-covariance matrix is $\mathbfss V(\bmath p)$, where the vectors $\btheta$ and
$\bmath p$ contain some free parameters to be estimated from the data. The corresponding
log-likelihood function may be expressed as
\begin{equation}
\ln \mathcal L(\btheta, \bmath p) = - \frac{1}{2} \ln \det \mathbfss V - \frac{1}{2}
{\bmath r}^{\mathrm T} \mathbfss V^{-1} \bmath r + N \ln \sqrt{2\pi},
\label{loglik}
\end{equation}
where $\bmath r(\btheta) = \bmath x-\bmu(\btheta)$ is the vector of the RV residuals. For
shortness, we will also denote the combined vector of all parameters $\btheta$ and $\bmath
p$ as $\bxi$.

Maximizing~(\ref{loglik}) over $\btheta$ and $\bmath p$, we obtain the best fitting
estimations of these parameters (the point where the maximum is attained). The value of
the likelihood maximum itself may further be used, for instance, in the likelihood ratio
test comparing two different data models.

As we explain in \citep{Baluev08b}, in practice it might be useful to replace the true
likelihood function~(\ref{loglik}) by a modified version
\begin{equation}
\ln \tilde \mathcal L(\btheta, \bmath p) = - \frac{1}{2} \ln \det \mathbfss V - \frac{1}{2\gamma}
{\bmath r}^{\mathrm T} \mathbfss V^{-1} \bmath r + N \ln \sqrt{2\pi},
\label{loglikmod}
\end{equation}
where the correction divisor $\gamma=1-\dim \btheta/N$. The goal of this modification is
to reduce the systematic bias that would otherwise appear in the noise parameters, because
the best-fit residuals $\bmath r$ are systematically smaller than real errors. This effect
is clear, e.g., in Fig.~\ref{fig_wj} that we will discuss in detail in a further section.
The bootstrap simulation in this plot (left panel) show clear systematic bias because the
bootstrap is based on the unscaled best-fit RV residuals. If not the
correction~(\ref{loglikmod}), the plain Monte Carlo simulations (right panel) would
demonstrate the same or simular bias, while the bias of the bootstrap would be effectively
doubled. Note that the modification~(\ref{loglikmod}) keeps intact all asymptotic
properties of the maximum-likelihood method; it only improves its behaviour when $N$ is
not so large.

Note that the generalized model~(\ref{loglik}) differs from the one used in
\citep{Baluev08b} in the matrix $\mathbfss V$, which is no longer diagonal. However the
general theory of maximum-likelihood estimations is basically the same for the both cases.
For example, to find the covariance matrix of the maximum-likelihood estimations, we
should first calculate the quadratic Taylor approximation of the function $\ln \mathcal
L(\bxi)$. From~(\ref{loglik}) we can easily derive the relevant gradient:
\begin{eqnarray}
\frac{\partial \ln \mathcal L}{\partial p_i} &=&
- \frac{1}{2} \trace \left( \mathbfss V^{-1} \frac{\partial \mathbfss V}{\partial p_i} \right)
+ \frac{1}{2} {\bmath r}^{\mathrm T} \mathbfss V^{-1} \frac{\partial \mathbfss V}{\partial p_i}
\mathbfss V^{-1} \bmath r, \nonumber\\
\frac{\partial \ln \mathcal L}{\partial\theta_i} &=&
{\bmath r}^{\mathrm T} \mathbfss V^{-1} \frac{\partial \bmu}{\partial\theta_i},
\label{loglik_gradient}
\end{eqnarray}
as well as the second-order derivatives:
\begin{eqnarray}
\frac{\partial^2 \ln \mathcal L}{\partial p_i \partial p_j} &=&
\frac{1}{2} \trace \left[ \mathbfss V^{-1} \left(\frac{\partial \mathbfss V}{\partial p_i}
\mathbfss V^{-1} \frac{\partial \mathbfss V}{\partial p_j} -
\frac{\partial^2 \mathbfss V}{\partial p_i \partial p_j} \right) \right] - \nonumber\\
& & - {\bmath r}^{\mathrm T} \mathbfss V^{-1} \left(
\frac{\partial \mathbfss V}{\partial p_i} \mathbfss V^{-1}
\frac{\partial \mathbfss V}{\partial p_j} -
\frac{1}{2} \frac{\partial^2 \mathbfss V}{\partial p_i \partial p_j}
\right) \mathbfss V^{-1} \bmath r, \nonumber\\
\frac{\partial^2 \ln \mathcal L}{\partial\theta_i \partial\theta_j} &=&
- \frac{\partial \bmu^{\mathrm T}}{\partial\theta_i} \mathbfss V^{-1} \frac{\partial \bmu}{\partial\theta_j} +
{\bmath r}^{\mathrm T} \mathbfss V^{-1} \frac{\partial^2 \bmu}{\partial\theta_i \partial\theta_j}, \nonumber\\
\frac{\partial^2 \ln \mathcal L}{\partial\theta_i \partial p_j} &=&
- {\bmath r}^{\mathrm T} \mathbfss V^{-1} \frac{\partial \mathbfss V}{\partial p_j}
\mathbfss V^{-1} \frac{\partial \bmu}{\partial\theta_i}.
\label{loglik_hesse}
\end{eqnarray}

Considering together~(\ref{loglik}-\ref{loglik_hesse}), we can write down the following
quadratic approximation:
\begin{equation}
\ln \mathcal L(\bxi) \simeq \ln \mathcal L(\hat\bxi) + \bmath g \cdot \bmath \Delta\bxi -
                            \frac{1}{2} \Delta\bxi^{\mathrm T} \mathbfss F \Delta\bxi,
\label{loglik_quadr}
\end{equation}
where $\Delta\bxi = \bxi-\hat\bxi$ with $\hat\bxi$ standing for the vector of the true
parameters, $\bmath g = \partial\ln\mathcal L/\partial\bxi$ is the compound gradient of
$\ln\mathcal L$ and $\mathbfss F$ is the Fisher information matrix:
\begin{equation}
\mathbfss F = \expect \left( \frac{\partial \ln \mathcal L}{\partial \bxi} \otimes
                             \frac{\partial \ln \mathcal L}{\partial \bxi} \right) =
            - \expect \frac{\partial^2 \ln \mathcal L}{\partial \bxi^2},
\label{fisher_def}
\end{equation}
where the expectation should be taken at the true parameter values. The elements of
$\mathbfss F$ in our case are
\begin{eqnarray}
F_{p_i p_j} &=& \frac{1}{2} \trace \left( \mathbfss V^{-1} \frac{\partial \mathbfss V}{\partial p_j}
                                          \mathbfss V^{-1} \frac{\partial \mathbfss V}{\partial p_i} \right), \nonumber\\
F_{\theta_i \theta_j} &=& \frac{\partial \bmu^{\mathrm T}}{\partial\theta_i} \mathbfss V^{-1} \frac{\partial \bmu}{\partial\theta_j}, \nonumber\\
F_{\theta_i p_j} &=& 0.
\label{fisher}
\end{eqnarray}

The expansion~(\ref{loglik_quadr}) allows us to approximate the point $\bxi^*$, where the
maximum is achieved, as $\bxi^* \simeq \hat\bxi + \mathbfss F^{-1} \bmath g$. Since the
relation $\var \bmath g = \mathbfss F$ holds true, the variance-covariane matrix of our
estimations $\bxi^*$ has the same asymptotic representation for large $N$ as in the
uncorrelated case:
\begin{equation}
\var \bxi^* \simeq \mathbfss F^{-1}.
\end{equation}
Notice that the vectors $\btheta^*$ and $\bmath p^*$ appear therefore asymptotically
uncorrelated, as in \citep{Baluev08b}, since the cross term $F_{\btheta \bmath p}$ is
again zero.

The numerical non-linear maximization of~(\ref{loglik}) or~(\ref{loglikmod}) can be
performed by means of the Levenberg-Marquardt algorithm. Notice that the simplified
widespread version of this algorithm that minimizes a sum-of-squares function (as
implemented, e.g., in the MINPACK package) is unsuitable here, because it relies on
certain relationships between the gradient and Hessian matrix, which are invalid in our
case. We need a general variant of the Levenberg-Marquardt algorithm \citep[e.g.][]{Bard}
that can maximize an arbitrary non-linear target function and can deal separately with the
gradient and the Hessian. Note that it is handy to approximate the Hessian as $-\mathbfss
F$ due to the expansion~(\ref{loglik_quadr}). It is useful because $\mathbfss F$ is always
positive definite. Besides, a lot of care is needed to optimize the calculational
performance, since the formulae~(\ref{loglik_gradient})-(\ref{fisher}) require very
computation-greedy operations with large matrices and vectors. We give some tips
concerning this issue in the Appendix~\ref{sec_speed}.

We only need to detail the last thing, namely the model of the noise covariance matrix
$\mathbfss V$. In this paper we consider three main noise models:
\begin{enumerate}
\item White-noise model. The matrix $\mathbfss V$ is diagonal; the diagonal elements
represent the total variances of individual RV measurements and are equal to the sum of
the instrumental part (the square of the stated measurement uncertainty) and the RV
jitter. The RV jitter is different for different instruments. This model was considered in
\citep{Baluev08b}.
\item Shared red-noise model. In addition to the white-noise components, we add to
$\mathbfss V$ the red-noise covariance matrix $\sigma_{\rm red}^2 \mathbfss R(\tau)$,
where the elements of $\mathbfss R$ are defined via some guessed noise correlation
function $\rho(x)$ as $R_{ij}(\tau)=\rho((t_j-t_i)/\tau)$. We chose $\rho(x)=e^{-|x|}$.
This noise model infers that the red noise belongs to the star, while the spectrographs
generate only the white noise.
\item Separated red-noise model. It is similar to the previous case, but the parameters
$\sigma_{\rm red}$ and $\tau$ are different for different instruments (HARPS and Keck).
The cross correlation between HARPS and Keck measurements is set to zero. This model
infers that the red noise belongs to the spectrographs, and not to the star.
\end{enumerate}

\section{GJ581 data analysis}
\label{sec_analysis}

\subsection{Preliminary investigation}
\label{subsec_validity}
The main goal of this subsection is to estimate the validity of various statistical
methods in the case of GJ581 RV data analysis. We have two datasets at our disposal: the
$240$ HARPS and $121$ Keck/HIRES RV measurements published in \citep{Forveille11} and
\citep{Vogt10}, respectively. First of all, we provide four-planet white-noise fit in
Table~\ref{tab_GJ581_bcde_white} that was obtained by means of the likelihood function
maximuzation as described in \citep{Baluev08c,Baluev08b,Baluev11}.

\begin{table*}
\caption{Best fitting parameters of the GJ581 planetary system: white jitter, four planets}
\label{tab_GJ581_bcde_white}
\begin{tabular}{@{}lllll@{}}
\hline
\multicolumn{5}{c}{planetary parameters} \\
\hline
                       & planet b         & planet c       & planet d         & planet e         \\
\hline
$P$~[day]              & $5.368585(79)$   & $12.9175(19)$  & $66.616(79)$     & $3.14922(18)$    \\
$\tilde K=K\sqrt{1-e^2}$~[m/s]
                       & $12.58(16)$      & $3.26(16)$     & $1.95(17)$       & $1.79(16)$       \\
$e$                    & $0.021(13)$      & $0.053(48)$    & $0.259(83)$      & $0.164(89)$      \\
$\omega$~[$^\circ$]    & $334(35)$        & $145(53)$      & $342(18)$        & $156(31)$        \\
$\lambda$~[$^\circ$]   & $142.93(76)$     & $106.5(3.0)$   & $144.9(5.2)$     & $63.3(5.4)$      \\
\hline
$M \sin i$~[$M_\oplus$]& $15.78(20)$      & $5.48(27)$     & $5.65(49)$       & $1.88(17)$       \\
$a$~[AU]               & $0.04061187(40)$ & $0.0729244(70)$& $0.21768(17)$    & $0.0284573(11)$  \\
\hline
\multicolumn{5}{c}{data series and common fit parameters} \\
\hline
                           & HARPS          & Keck            & & \\
$c_0$~[m/s]                & $-9205.96(13)$ &  $1.08(27)$     & & \\
$\sigma_{\rm white}$~[m/s] & $1.50(11)$     &  $2.45(23)$     & & \\
\hline
r.m.s.~[m/s]               & $1.96$         &  $2.82$         & & \\
$\tilde l$~[m/s]           & \multicolumn{2}{c}{$2.25$}       & & \\
\hline
\end{tabular}\\
The $M \sin i$ and $a$ values were derived assuming the mass of the star $M_\star = 0.31
M_\odot$, which was used e.g. by \citet{Forveille11}. The uncertainty of $M_\star$ were
not included in the uncertainties of the derived values. The mean longitudes $\lambda$
refer to the epoch $JD2454500$. The goodness of the fit $\tilde l$ was derived from the
maximum of $\ln \mathcal L$ as explained in \citep{Baluev08b}.
\end{table*}

The maximum-likelihood approach infers a set of well-known classical theoretical results
and methods concerning the maximum-likelihood estimations, that were established under a
condition that the number of observations tends to infinity. We will call them
collectively as Asymptotic Maximum-Likelihood Estimation Theory (hereafter AMLET). A
fraction of them is described in \citep{Baluev08b}, bearing in mind an application to the
exoplanetary RV curve fitting task.

Notice that AMLET tools are sometimes called as frequentist ones, especially in the works
employing the Bayesian analysis, where such opposing highlights the advantages of the
Bayesian methods. Such terminology actually hides a misconception: AMLET represents,
basically, a common limit to which both frequentist and Bayesian approaches converge, when
the number of observation tends to infinity. Therefore, it would be incorrect to equate
AMLET and the general frequentist approach in the statistics, since most of the classical
AMLET propositions can be easily reinterpreted from the Bayesian point of view, while the
genuine frequentist methods in their general form \citep[e.g.][]{Lehman} are more
complicated and theoretically justified than AMLET.

It is often argued that AMLET tools are often not applicable to the exoplanetary RV data
analysis, especially for multi-planet systems which involve very complicated non-linear RV
signal models. However, this is rarely verified with concrete practical cases. In this
paper we undertook an attempt to rigorously assess the applicability of AMLET tools to the
case of the GJ581. Since this research invloves a vast amount of various numerical
simulations and rather boring statistical stuff, which is not related directly to the
GJ581 system itself, we do not describe these results in the main body of the paper. An
interested reader may find the detailed discussion in the Appendix~\ref{sec_AMLET}. Here
we only provide a short summary of our investigation:
\begin{enumerate}
\item In the case of the GJ581 RV data, the parametric confidence regions and false
alarm probabilities, obtained using AMLET, work well for the white-noise and shared
red-noise 4-planet models, but are unsuitable for the separated red-noise model. This
indicates that the latter model is over-parametrized and must be used with caution.

\item When analysing the HARPS and Keck data independently from each other, we may use
AMLET for the HARPS time series, but not for the Keck one. Actually, the Keck dataset is
the main thing that makes AMLET unusable with the separated red-noise RV model.

\item We should avoid using the bootstrap simulation (section~\ref{subsec_bstrp}) for any
of our models, because it works in an unexpected and misleading manner when a
parameterized noise is involved. However, we may safely use the usual Monte Carlo
simulation (section~\ref{subsec_MC}) instead, since the RV data show absolutely no hints
of any non-Gaussianity, which is the main fear of people prefering the
bootstrap.\footnote{Sometimes it is claimed that a correlated noise is not Gaussian. We
must caution the reader against such mixing of distinct notions. In our case of GJ581, for
instance, the noise is consistent with a Gaussian random process. Such a process has
Gaussian individual values, which are nonetheless mutually correlated.}

\item In the most cases, we have no need for complicated and computationally greedy
techniques like the Bayesian analysis or the genuine frequentist methods. For the
white-noise and shared red-noise model we can just use AMLET with no fear. For the
separated red-noise model AMLET is poor, but with this model we obtain little serious
results that would need a deep verification.
\end{enumerate}

\subsection{Assessing the significance of the red noise}
Let us first assess rigorously the significance of the noise non-whiteness. We can do this
using the method described in \citep{Baluev11}. Using a variation of the Monte Carlo
algorithm~\ref{subsec_MC} from the Appendix~\ref{sec_simul}, we generated a bunch of
$1000$ simulated residual periodograms assuming the white model of the noise and 4-planet
model of the RV curve. Thus, these periodograms are evaluated using exactly the same
algorithm as in Fig.~\ref{fig_rnGJ581}, but based on simulated uncorrelated data. Each
simulated periodogram is then smoothed, also exactly as in Fig.~\ref{fig_rnGJ581}. Based
on this set we can derive the distribution of single values of the smoothed periodograms
(for an arbitrary frequency), and also the distribution of the associated $\max/\min$
ratio, which characterizes the degree of non-whiteness of the simulated spectrum. It is
important that this procedure does not require us to make any assumptions about the
red-noise correlation function.

\begin{figure}
\includegraphics[width=84mm]{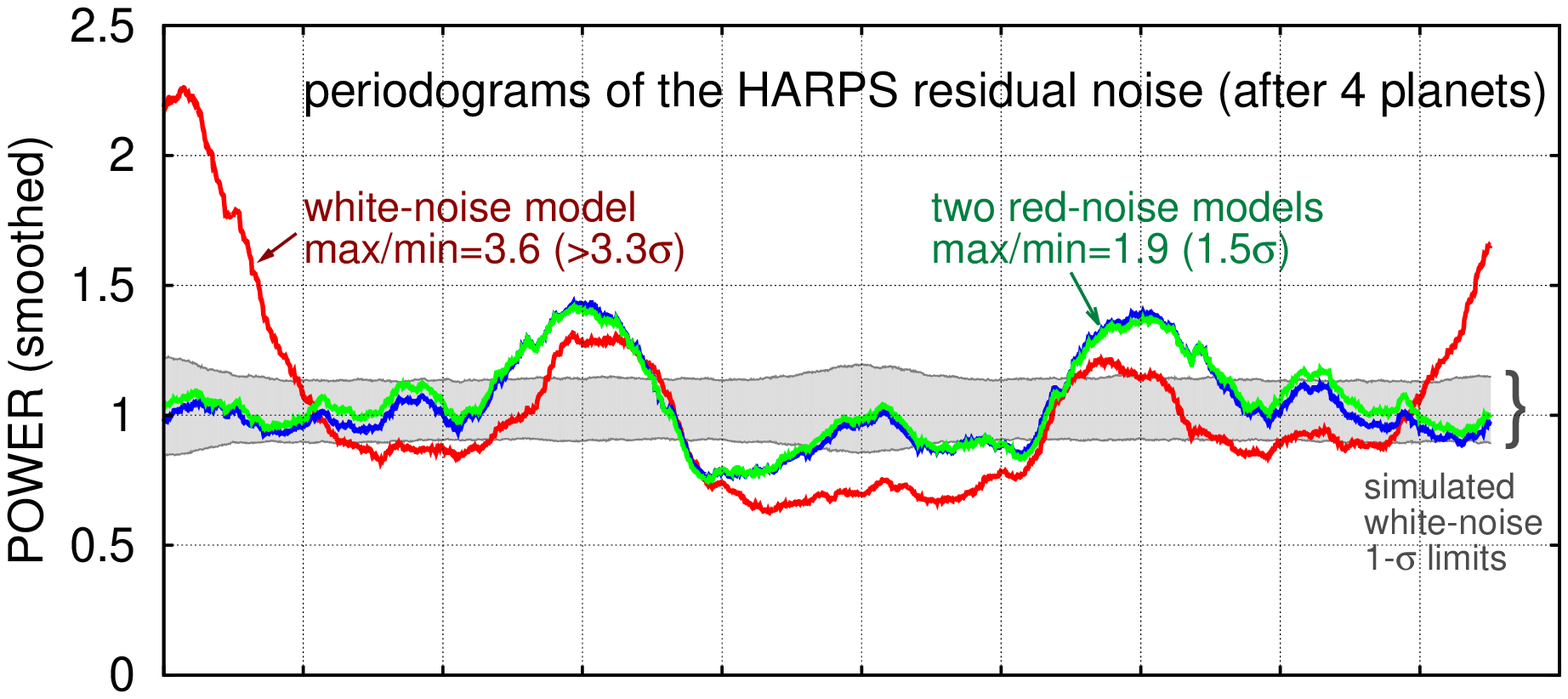}
\includegraphics[width=84mm]{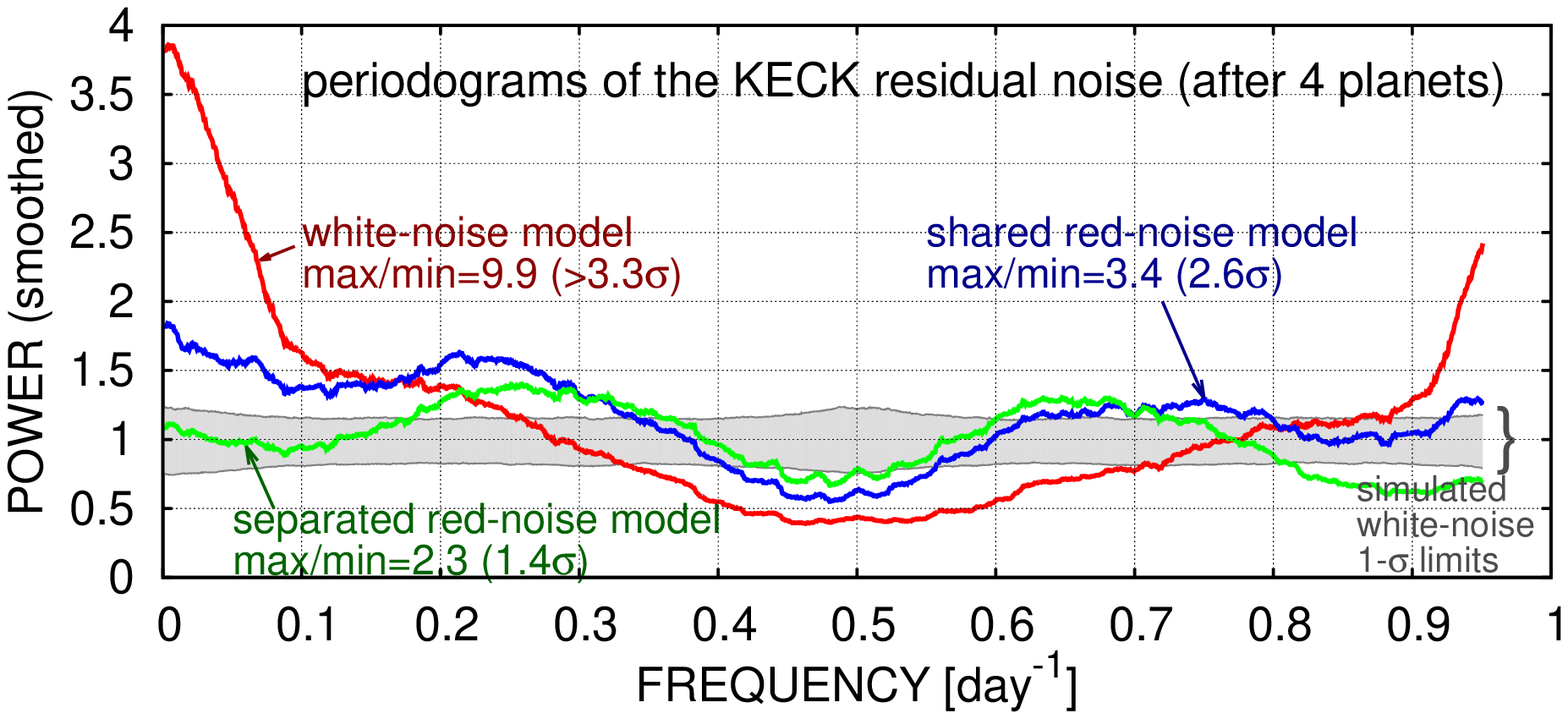}
\caption{The periodograms of the HARPS and Keck residual noise constructed for various
noise models, in comparison with simulated limits expected for the white-noise case. We
can see that periodograms based on the white-noise model show large variations with
$\max/\min$ ratio well above the $3\sigma$ level, and using the red-noise models
significantly suppresses these variations.}
\label{fig_rnsigGJ581}
\end{figure}

The results are shown in Fig.~\ref{fig_rnsigGJ581}. We can see that among $1000$ Monte
Carlo trials none could reproduce the same large $\max/\min$ ratios that we observed for
the smoothed periodogram of the real data. Therefore, the non-whiteness in the RV noise of
these real data has very high significance ($>99.9\%$). For comparison, we also plot the
periodograms of the real data on the basis of the red-noise models, utilising the
algorithm of Section~\ref{sec_reduct}. We can see that these frequency spectra are already
consistent with the white-noise statistical limits, possibly except for the case of the
Keck periodogram with shared red-noise model, where the residual non-whiteness has the
significance of $2.6\sigma$. Therefore, this model may be incapable of complete
elimination of the red noise, probably because the red noise has somewhat different
characteristics between the HARPS and Keck datasets. We think, however, that this shared
red-noise model suits our practical needs at best, since the residual frequency spectrum
non-whiteness is anyway a few times smaller than it was for the original white-noise
model. The separated red-noise noise model can do apparently more impressive reduction of
the correlated noise, but, as we have discussed above, this model is statistically poor.

We must emphasize that all significance levels printed in Fig.~\ref{fig_rnsigGJ581},
including the one of $2.6\sigma$ for the Keck shared red-noise periodogram, were derived
from white-noise simulations. To be fully honest, we ought to evaluate these levels
assuming a matching noise model for each, but it appeared too computationally-demanding
for the red-noise cases. We expect that the correct significances for red-noise
periodograms may be somewhat smaller, because such periodograms usually showed
systematically higher significance levels. It may even appear, that the mentioned residual
Keck non-whiteness for the shared red-noise case is eventually insignificant. However,
this does not alter our conclusion that the white-noise model is inadequate; the relevant
significance is based on the correct (matching) noise model and is well above the
$3\sigma$ level, both for the HARPS and Keck data.

\subsection{Detailed analysis of the RV data}
\subsubsection{HARPS data alone}
\label{sss_HARPS}
\begin{figure*}
\includegraphics[width=0.85\textwidth]{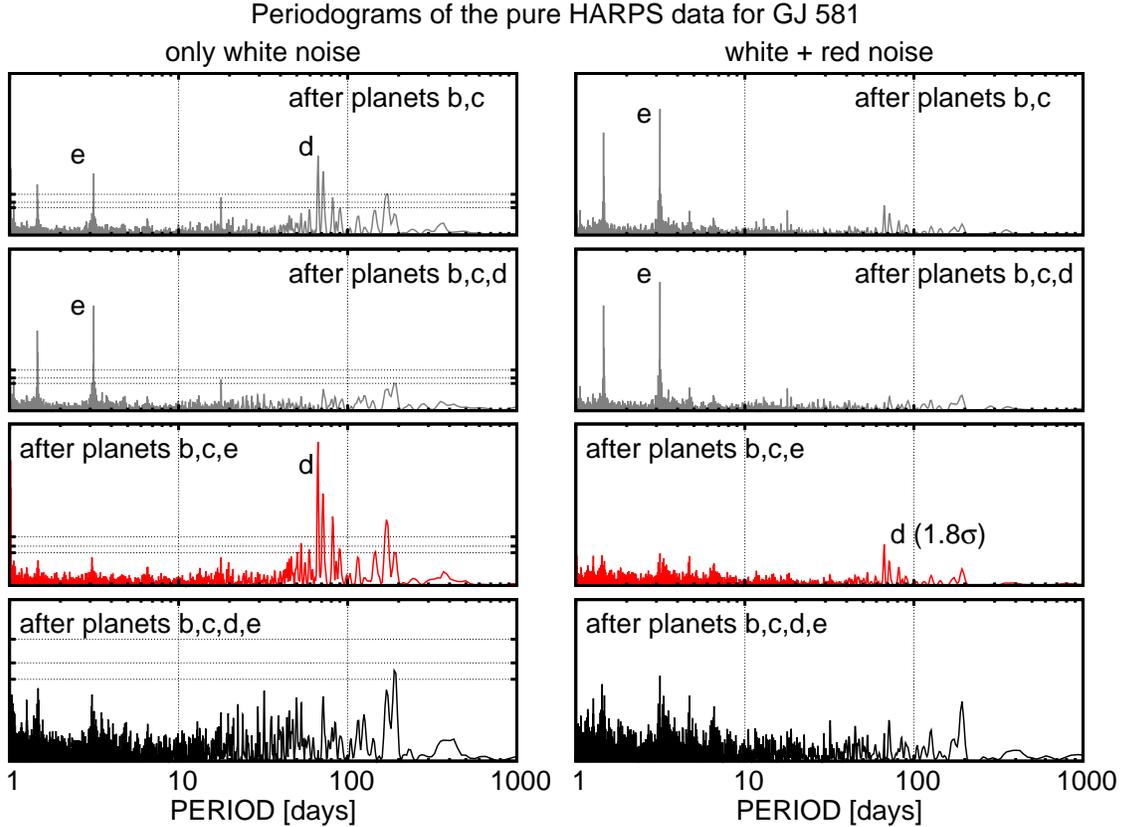}
\caption{Residual periodograms of the HARPS RV data of GJ581 constructed on the basis of
various base RV curve models. The base models include the compound multi-Keplerian signal
from the planets labelled in a relevant plot. Note that the plots in the same row always
have the same ordinate range. In the white-noise plots we show the $1$-, $2$-, and
$3$-$\sigma$ significance levels derived in accordance with \citep{Baluev08a}. These
levels are in good agreement with Monte Carlo simulations. The technique from
\citep{Baluev08a} was not designed to work with correlated data, so for the red-noise
plots we only show the simulated significance of the most interesting planet \emph{d} peak
(third periodogram in the red-noise column).}
\label{fig_GJ581_HARPS}
\end{figure*}

Let us start from the analysis of $240$ HARPS RV measurements. In
Fig.~\ref{fig_GJ581_HARPS} we show a series of the residual periodograms, starting from
the two-planet base model (planets \emph{b} and \emph{c}). In the case of the white-noise
model we are able to subsequently extract all four planets from these periodograms. We can
see that all four peaks show very high significance. However, in the last residual
periodogram, corresponding to the case when all four peaks are already extracted, we can
see a typical red noise picture: an amorphous set of peaks at the periods longer than
$\sim 10$~d, a diurnal alias of this frequency band close to $1$~d period, and a
depression in the middle part of the period range.

We can see that our maximum-likelihood algorithm suppressed the effect of the red noise,
as expected. However, together with the red noise, our procedure dramatically suppressed
the planet \emph{d} peak at $67$~d. This is not very surprising on itself, since this peak
is in the range where the red noise is ruling. However, the final significance of this
peak becomes marginal~-- only $1.8\sigma$~-- making us rather sceptical about the reality
of this planet.

Speaking shortly, although we cannot claim that the planet \emph{d} RV signature is
insignificant, we must admit that its detection is not robust and requires a serious
verification. The relevant RV variation may be caused by correlated RV noise in the data,
and does not necessarily reflect a Doppler wobble induced by a real planet.

\subsubsection{Keck data alone}
\label{sss_KECK}
\begin{figure*}
\includegraphics[width=0.85\textwidth]{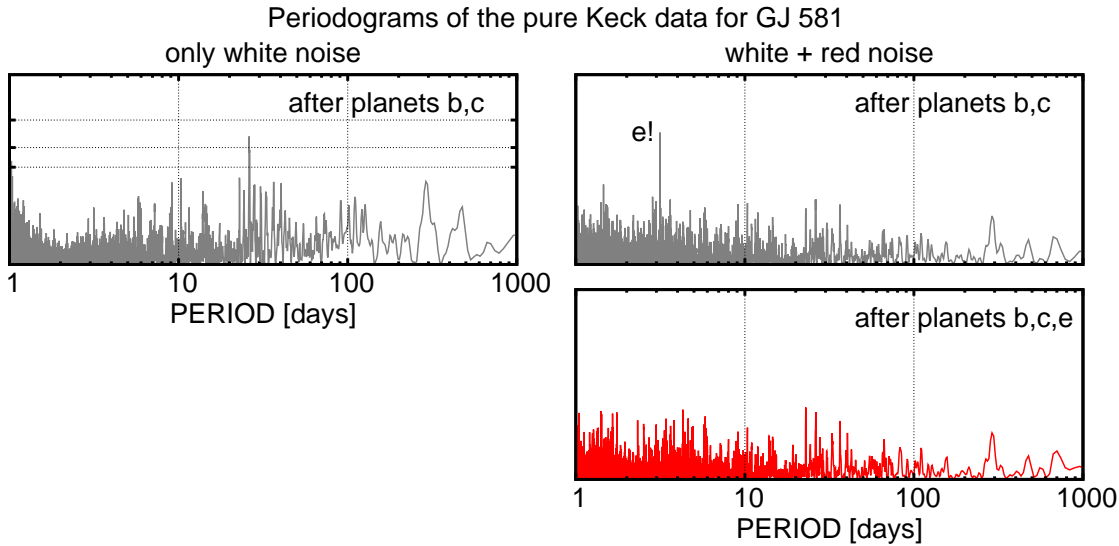}
\caption{Same as in Fig.~\ref{fig_GJ581_HARPS}, but for the Keck dataset. We can see that
the white-noise model does not allow robust detection of any real planet except for the
planets \emph{b} and \emph{c} (which both are undoubtful anyway), but applying the
red-noise model allows to confirm the existence of the planet \emph{e} (see text for the
detailed calculation of its significance). No hints of the planet \emph{d} can be revealed
with any of these models, however.}
\label{fig_GJ581_KECK}
\end{figure*}

Let us now deal with $121$ Keck/HIRES RV measurements in the similar manner. The relevant
periodograms are shown in Fig.~\ref{fig_GJ581_KECK}. First, we can see that now we cannot
detect more than two planets \emph{b} and \emph{c}, if we use the traditional white-noise
model. This conclusion is in the agreement with previous studies
\citep{Gregory11,dosSantos12}, but it is still rather disappointing, because the Keck RV
precision is pretty competitive in comparison with the HARPS one. Second, in the Keck data
we again see clear hints of the red noise, which created a fake periodicity at
approximately $27$~days. In this case our red-noise removing algorithm does its job even
better than anyone could expect: it did not just killed all fake red-noise peaks, but it
also reveals the $3.1$-day peak belonging to the planet \emph{e}! This proves that our
algorithm does not just suppresses the apparent RV variations, lifting up the detection
thresholds. It is working in a much more intelligent manner: in certain frequency ranges
it may basically \emph{improve} the effective RV precision, revealing the true
periodicities that the red noise tries to hide.

The period of the newly discovered variation in the Keck data is in excellent agreement
with the planet \emph{e} period obtained from the HARPS data. Such coincidence is hardly
casual. However, what is its rigorous significance? The answer to this question is not
obvious, because we cannot use AMLET for the Keck dataset alone. Monte Carlo simulations
(algorithm \ref{subsec_MC}) suggest that the significance associated to this peak of the
Keck periodogram is only $1.2\sigma$. Therefore, if we tried to \emph{detect} this planet
from the Keck data alone, with absolutely no reference to the HARPS data, we would have to
admit that this peak is statistically insignificant. Basically, it is a luck that no other
comparable peak appeared in the top-right periodogram of Fig.~\ref{fig_GJ581_KECK}.

However, we need just to \emph{confirm} the planet \emph{e} existence on the basis of the
Keck dataset, not to \emph{detect} it anew. This places much more mild limits. We have no
need to simulate the periodogram in its whole period range as shown in
Fig.~\ref{fig_GJ581_KECK}. We already know the probable planet \emph{e} parameters from
the HARPS data with good precision, including e.g. its orbital period. Now we only need to
confirm that RV noise could not generate so large peak as we can see in the Keck data just
\emph{in a narrow vicinity} around this known period. It is not a big deal if we find a
noisy peak at some faraway frequency, where the real planet \emph{e} definitely cannot
reside. Such \emph{confirmational} significance will be much larger than the
\emph{detectional} one, because the probability for the RV noise to occasionally generate
a large peak inside a narrow frequency segment is much smaller than inside a wide one.
This becomes obvious if we look at the periodogram's false-alarm probability approximation
from \citep{Baluev08a}:
\begin{equation}
\FAP \lesssim W e^{-z} \sqrt z.
\end{equation}
We can see that this estimation depends on the normalized frequency bandwidth $W$. For the
whole range of periods from $1$~day to infinity we have $W\approx 3500$ (for the Keck
dataset taken alone), but when dealing with confirmational false alarm probabilities, we
need to consider $W\sim 1$ at most, since the $\pm 1\sigma$ uncertainty range of the
expected planet \emph{e} period corresponds to $W\sim 0.1$. Therefore, the confirmational
false alarm probability might be a few thousand times smaller that the detectional one.

However, we must remember that the red-noise model infers strong non-linearity when used
with Keck data alone, as we have already discussed above. We cannot use any asymptotic
methods in this case, and our numerical simulations must be more intricate than the plain
Monte Carlo scheme \ref{subsec_MC}. We can no longer rely on a single simulation series
based on a single vector of nominal ``true'' model parameters, as we have done before.
Instead, we should honour some representative parametric domain. Since the true values may
be anywhere in this domain, we must generate many distinct simulation series according to
the Monte Carlo scheme from Section~\ref{subsec_MC}, each time assuming different vector
for the mock ``true'' parameters. The detailed step-by-step description of this algorithm
is given in Section~\ref{subsec_FMC}.

\begin{figure*}
\includegraphics[width=0.85\textwidth]{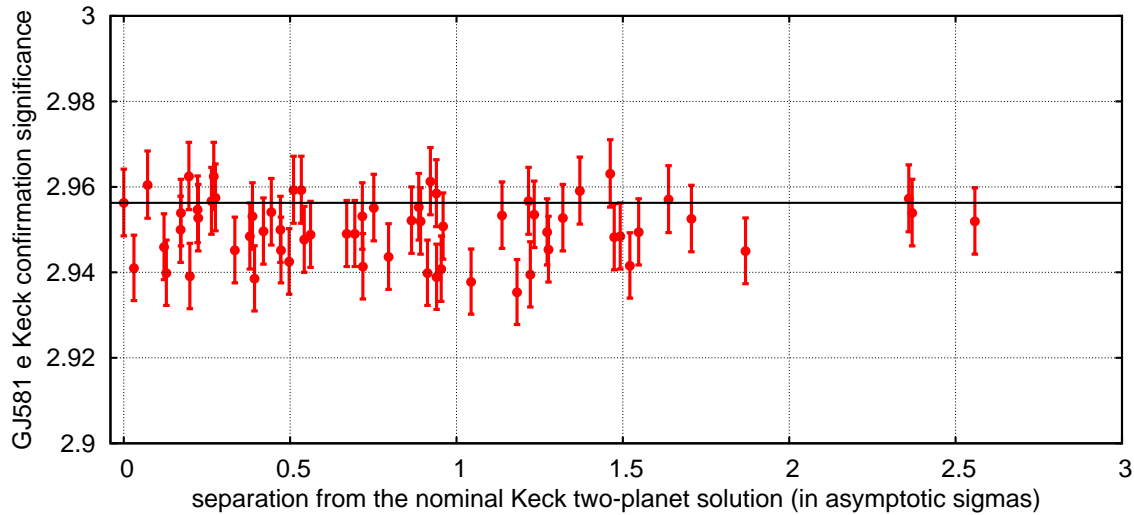}
\caption{Simulated confirmation significance of the planet GJ581~\emph{e}, depending on
the assumed true parameters of the two-planet red-noise model of the Keck RV data. The
graph shows a set of points, each corresponding to a different test ``true'' vector of
parameters, simulated according to the Monte Carlo algorithm~\ref{subsec_MC}. The
abscissas of the points reflect the deviation of these trial ``true'' parameters from the
nominal best-fit ones, expressed in terms of the asymptotic $n$-$\sigma$ significance
level, which was derived from the relevant value of $\tilde Z$. The ordinate of each point
shows the confirmation significance of the planet \emph{e}, as derived from $500000$ Monte
Carlo simulations of the relevant Keck periodogram (see text for details). The errorbars
reflect the statistical uncertainties inferred by these second-level Monte Carlo
simulations. We can see that the confirmation significance almost reaches the $3\sigma$
level and is practically independent on the assumed two-planet configuration.}
\label{fig_e_sig}
\end{figure*}

During this calculation, we first generated a sequence of the trial ``true'' sets of
parameters, assuming the two-planet red-noise Keck RV data model (the first-level
simulation). This first-level simulation is not intended to have big statistical meaning,
we need it just to obtain a set of points covering some more or less wide domain around
the best fitting two-planet solution. After that, for each of the generated parametric
vectors, we run the plain Monte Carlo simulation (algorithm \ref{subsec_MC}, $500000$
random trials in each simulation). On each random trial of this second-level simulation we
generate an artificial Keck dataset using the model ``two-planet RV variation + correlated
RV noise'' (without planet \emph{e}). Then for each such dataset we evaluate the relevant
residual periodogram exactly in the same way as in the top-right panel of
Fig.~\ref{fig_GJ581_KECK}, where we used the real Keck data. For each such periodogram we
find the maximum in the narrow period range $3.145-3.153$~day ($W\approx 3$, centered at
the nominal planet \emph{e} period). Based on such Monte Carlo sequence, we count the
fraction of simulated periodograms that demonstrated the same or larger maximum peak as
the one that we have seen for the real data. This fraction represents the desired
confirmational false-alarm probability of the planet \emph{e}, as inferred by the adopted
``true'' two-planet model. These false-alarm probabilities can be further transformed to
the normal (``$n$-$\sigma$'') significance levels that we use throughout the paper.

We plot the results of these simulations in Fig.~\ref{fig_e_sig}. From this graph, we can
see that the simulated confirmational significance practically does not depend on the
adopted parameters of the base two-planet model, even when these parameters deviate from
the nominal estimation by more than $2\sigma$. Actually, most of the scatter around the
nominal level is likely due to the statistical uncertainty of the second-level Monte
Carlo. If we generated more trials for each point in Fig.~\ref{fig_e_sig}, this scatter
would probably shrink further. Basically, this figure does not reveal any real dependence
on the true parameters (at least in the parametric domain that we were able to fill in the
first-level Monte Carlo). The rigorous frequentist significance level is given by the
minimum ordinate among all simulated points, while the nominal level corresponds to the
point located at zero abscissa. These values do not differ much and can be rounded to
$3\sigma$ both. This means that Keck data can robustly confirm the existence of the planet
GJ581~\emph{e} RV signal.

Our algorithm does not reveal any hint of the planet \emph{d} in the Keck data. The
corresponding residual periodogram calculated after extraction of the three planets
\emph{b}, \emph{c}, and \emph{e}, looks like a perfect white noise with no peak attracting
any attention. Maybe this planet \emph{d} does not actually exist, and the variation that
we have seen in the HARPS data is some systematic effect or just some random fluctuation?
Unfortunately, Keck data alone cannot supply an independent answer to this question.
Notice that there is a pretty large difference in the significance of the planet
\emph{e}, as inferred by the HARPS and Keck data. After extrapolation of this difference
to the planet \emph{d}, we realize that currently available Keck data are just unabe to
reveal it, no matter exists it or not.

\subsubsection{Combined dataset}
\label{sss_joint}
Now let us proceed to the joint analysis of the HARPS and Keck data. We consider three
noise models that we have already introduced: the white-noise model, the shared red-noise
model, and the separated red-noise model.

\begin{figure*}
\includegraphics[width=0.99\textwidth]{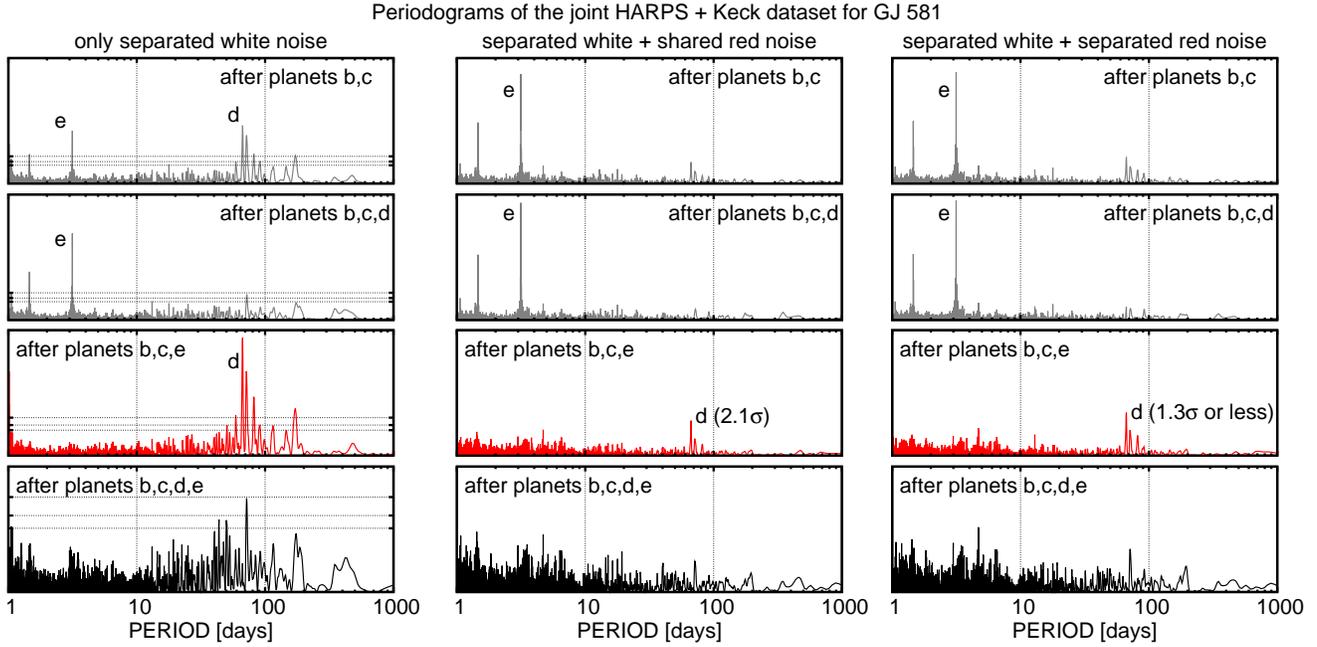}
\caption{Same as in Fig.~\ref{fig_GJ581_HARPS}, but for the combined HARPS+Keck dataset.}
\label{fig_GJ581_joint}
\end{figure*}

We show a series of the relevant periodograms in Fig.~\ref{fig_GJ581_joint}. We can see
that while the planets \emph{b}, \emph{c}, and \emph{e} can be robustly extracted from
these data, the planet \emph{d} still remains rather controversial, because its
significance drops to only $\sim 2.1\sigma$ or even below, if a red-noise model is used.
We feel such significance level is not enough to claim a robust detection of an exoplanet,
because this significance is model-dependent. The planet candidate GJ581~\emph{d} should
be probably reclassified as a controversial one.

Finally, we present two fits of the GJ581 planetary system, obtained using the shared
red-noise model. In the first fit (Table~\ref{tab_GJ581_bce}) we provide a three-planet
configuration with planets \emph{b}, \emph{c}, and \emph{e}, while the second one
(Table~\ref{tab_GJ581_bcde}) also involves planet \emph{d}.

\begin{table*}
\caption{Best fitting parameters of the GJ581 planetary system: shared red jitter, three planets}
\label{tab_GJ581_bce}
\begin{tabular}{@{}llll@{}}
\hline
\multicolumn{4}{c}{planetary parameters} \\
\hline
                       & planet b         & planet c       & planet e         \\
\hline
$P$~[day]              & $5.368589(68)$   & $12.9186(21)$  & $3.14905(16)$    \\
$\tilde K=K\sqrt{1-e^2}$~[m/s]       & $12.65(13)$      & $3.20(17)$     & $1.69(13)$       \\
$e$                    & $0.022(10)$      & $0.040(44)$    & $0.195(73)$      \\
$\omega$~[$^\circ$]    & $38(25)$         & $122(61)$      & $38(24)$         \\
$\lambda$~[$^\circ$]   & $142.89(64)$     & $102.5(3.4)$   & $62.2(4.6)$      \\
\hline
$M \sin i$~[$M_\oplus$]& $15.86(16)$      & $5.38(28)$     & $1.77(13)$       \\
$a$~[AU]               & $0.04061189(34)$ & $0.0729286(80)$& $0.02845621(95)$ \\
\hline
\multicolumn{4}{c}{data series and common fit parameters} \\
\hline
                           & HARPS          & Keck            & \\
$c_0$~[m/s]                & $-9206.01(28)$ &  $0.86(34)$     & \\
$\sigma_{\rm white}$~[m/s] & $0.33(40)$     &  $1.18(28)$     & \\
$\sigma_{\rm red}$~[m/s]   & \multicolumn{2}{c}{$2.05(19)$}   & \\
$\tau_{\rm red}$~[day]     & \multicolumn{2}{c}{$11.0(3.4)$}  & \\
\hline
r.m.s.~[m/s]               & $2.43$         & $2.96$          & \\
$\tilde l$~[m/s]           & \multicolumn{2}{c}{$2.10$}       & \\
\hline
\end{tabular}\\
Same notes as in Table~\ref{tab_GJ581_bcde_white}.
\end{table*}

\begin{table*}
\caption{Best fitting parameters of the GJ581 planetary system: shared red jitter, four planets}
\label{tab_GJ581_bcde}
\begin{tabular}{@{}lllll@{}}
\hline
\multicolumn{5}{c}{planetary parameters} \\
\hline
                       & planet b         & planet c       & planet d         & planet e         \\
\hline
$P$~[day]              & $5.368603(66)$   & $12.9198(20)$  & $66.56(12)$      & $3.14905(15)$    \\
$\tilde K=K\sqrt{1-e^2}$~[m/s]       & $12.62(13)$      & $3.18(16)$     & $1.81(28)$       & $1.73(13)$       \\
$e$                    & $0.022(10)$      & $0.039(43)$    & $0.28(11)$       & $0.167(71)$      \\
$\omega$~[$^\circ$]    & $32(26)$         & $138(62)$      & $329(26)$        & $41(26)$         \\
$\lambda$~[$^\circ$]   & $143.06(63)$     & $105.4(3.2)$   & $143.0(8.8)$     & $62.5(4.5)$      \\
\hline
$M \sin i$~[$M_\oplus$]& $15.83(16)$      & $5.35(26)$     & $5.247(80)$      & $1.81(13)$       \\
$a$~[AU]               & $0.04061196(33)$ & $0.0729331(75)$& $0.21754(25)$    & $0.02845622(93)$ \\
\hline
\multicolumn{5}{c}{data series and common fit parameters} \\
\hline
                           & HARPS          & Keck            & & \\
$c_0$~[m/s]                & $-9205.96(22)$ &  $0.79(31)$     & & \\
$\sigma_{\rm white}$~[m/s] & $0.50(27)$     &  $1.24(26)$     & & \\
$\sigma_{\rm red}$~[m/s]   & \multicolumn{2}{c}{$1.65(16)$}   & & \\
$\tau_{\rm red}$~[day]     & \multicolumn{2}{c}{$9.3(3.1)$}   & & \\
\hline
r.m.s.~[m/s]               & $1.99$         & $2.79$          & & \\
$\tilde l$~[m/s]           & \multicolumn{2}{c}{$2.01$}       & & \\
\hline
\end{tabular}\\
Same notes as in Table~\ref{tab_GJ581_bcde_white}.
\end{table*}

\section{Reality of the putative fifth and sixth planets}
\label{sec_planet56}
Various authors have already raised a lot of doubts concerning the existence of the
planets GJ581 \emph{f} and GJ581 \emph{g}, since their announcement in \citep{Vogt10}.
Basically, it appears that their parameters may change significantly with each data update
and, besides, they do not demonstrate enough resistance with respect to various rather
subjective choices: methods of data analysis, model details, etc. For example, even for
the traditional white-noise model, we do not see in our work any hints of these planets,
as they were originally reported in \citep{Vogt10}. The four-planet residual periodograms
plotted here separately for the HARPS, Keck, and joint datasets demonstrate different
patterns of individual marginally significant maximum peaks, although they all reveal a
similar average power excess at the periods longer than $10$~days.

Our advanced analysis suggests that the red-noise models leave no room for any RV
variations beyond the four-planet models: neither in the HARPS, nor in the Keck, nor in
the combined dataset. Our red-noise models just absorbed all RV signals interpreted
previously as the hints of the planets \emph{f} and \emph{g}. However, maybe it is just a
question of interpretation? There are a lot of resons explaining why the measuments taken
at different epoch may be statistically correlated with each other. The genuine RV noise
caused by some astrophysical activity of the star, for instance, is a likely explanation,
but other sources are still possible. For example, a sinusoidal periodic oscillation would
produce the same periodic contribution in the compound correlation function of the data.
This might produce a picture very similar to what we see e.g. in Fig.~\ref{fig_corrGJ581}
until we extend this graph to larger time lags (which appears practically impossible
because the necessary regularity of the RV data is lost at large time separations). Then
why this red noise cannot be solely induced by some extra planets? Indeed, when working
only in the time domain (correlation functions) it may appear practically impossible to
disentangle the red noise from deterministic long- and moderate-term variations. However,
in our work we rely on the frequency domain (power spectra), where this task is not that
hard.

Of course, it remains theoretically possibile that this red noise represents just a
mixture of many \emph{true} periodic variations. Actually, the same logic can be applied
to the usual white noise equally well. For practice, this interpretation really changes
nothing: we still unable to model these hypothetical variations separately and we have to
find some compound model for them. It is the case where we should just apply the Occam
razor. The really meaningful question is: can we efficiently eleminate the red noise by
means of only a few periodic harmonics? We find that in the case of GJ581 the answer is
no. We started to honestly select the highest peaks from the last white-noise residual
periodogram in Fig.~\ref{fig_GJ581_joint} and subsequently remove the relevant
periodicities from the residuals, one by one, each time plotting a new residual
periodogram. We stopped after two such extra periodicities, because we did not achieve any
impressive progress (the power excesses for periods more than $10$~days and around the
diurnal period were still their), and the residual periodogram contained already
\emph{three} moderate peaks at the periods of a few tens of days. They had the same
marginal formal significance as the two previous ones. Clearly, only the red-noise models
allow to purge out all these chameleonic peaks at once.

Based on our investigation, we conclude that the putative planets GJ581~\emph{f} and
GJ581~\emph{g} likely do not exist, and the relevant RV signatures belong to the
correlated noise. We do not say that we reject the existence of absolutely any planet in
this system beyond the four known ones. However, even if some more planets exist in this
system, they are not detectable in the present data, and they are unrelated to the peaks
that we can see in the periodograms plotted with the use of only the white-noise model.
The current RV data really support existence of no more than four planets orbiting GJ581.

\section{Conclusions and discussion}
\label{sec_conclusions}
Although this work was focused on the concrete exoplanetary system of GJ581, we believe
that our results may have much more general meaning. The cases of GJ876 discussed in
\citep{Baluev11} and of GJ581 are not unique, and it seems that there are more examples of
planet-hosting stars demonstrating clear signs of the correlated noise in their publicly
available RV data. Actually, we believe that the red RV noise might be rather common
phenomenon.

This imposes bad as well as good concequences. The bad thing is that we have to use more
complicated and computationally slow methods of the analysis. Without that any analysis of
such data cannot be reliable. Unfortuanately, the functional shape of the correlation
function have not yet been investigated well, so we have to make some rather voluntaristic
guesses about it. Also, we need to accumulate rather large RV time series before the noise
correlation parameters become fittable. In the case of GJ581, for instance, the size of
the HARPS dataset is large enough, while the Keck data (half of the HARPS data in number)
are not so good in this concern.

The good thing is that the method of the red-noise modeling does not just suppress the
\emph{phantomic} RV variations together with anything else on its way; it is capable to
reveal \emph{true} variations that were hidden beyond the fog of correlated noise. This
means that our approach allows to increase, basically, the effective precision of the RV
measurements, at least in the short-period domain. This offers a way to partly overcome
the barrier set by the intrinsic RV jitter of the star, at least for some stars and in
certain frequency ranges. In particular, we belive that our method may decrease
exoplanetary detection threshold for active and/or subgiant stars, where the RV jitter
contribution dominates in the total error budget, making it impossible to obtain the RV
precision of better than $10-100$~m/s.

It is not yet fully clear, what is the source of the RV noise correlation. It may be
caused by some long-living spots or other details on the star's visible surface, or may be
a result of aggregation of various instrumental effects unrelated to the star itself. Our
statistical analysis cannot offer a definite answer to this question. We believe that in
general both reasons may be responsible. However, in the particular cases of GJ581 and
GJ876 the first interpretation seems more likely, because in both these cases the red RV
noise was detected in \emph{two} independent datasets (HARPS and Keck).

In view of this topic, we cannot leave aside a very recent report on the detection of a
``hot Earth'' orbiting $\alpha$~Cen B \citep{Dumusque12}. This discovery would not be
possible without careful elimination of various effects of stellar activity polluting the
periodograms at low frequencies. Basically, this team also applied some method of red
noise modeling, based on its correlation with a spectral activity index. Although this
method is different from our approach (we just unable to use the approach of
\citep{Dumusque12}, because all what we have is the public radial velocities of GJ581 and
not its raw spectra), we may note that the situation with the planet of $\alpha$~Cen~B
looks quite similar to the case of GJ581~\emph{e}. Anyway, the work by \citet{Dumusque12}
puts more emphasis of our conclusion that the future of the Doppler planet searches lies
in the careful treatment of the measured RV noise.

What concerns the particular case of the GJ581 planetary system, we were able to obtain
several important results. First, we have shown that its RV data do not really support
existence of any extra planet beyond the four-planet model. All apparent periodicities in
these data, that were previously interpreted as extra planets, are illusions caused by RV
noise correlations. Moreover, even the planet GJ581~\emph{d} becomes doubtful. Its
significance in the HARPS data does not even reach $2\sigma$, and in the Keck data it is
not detectable at all. In the combined dataset it may reach a more honourable level of
$2.1\sigma$ but this level is still model-dependent. We admit that this planet is not
rejected in our work and still remains rather probable, but despite of this fact we insist
that it should be reclassified as a controversial one, until more data (preferably
independent on HARPS) confirm it. The two-sigma significance level is not enough to claim
a robust planet detection, if it was not confirmed by independent observations. On
contrary, we were able to robustly confirm the planet GJ581~\emph{e}. This planet can be
revealed in the HARPS and Keck RV data independently, although to find its signal in the
Keck time series it is mandatory to use a red-noise data model. The confirmation
significance of this planet in the Keck data is $\sim 3\sigma$, although the same
detection significance is only $\sim 1\sigma$.

At last, we would like to discuss briefly the results concerning the GJ581 planetary
system presented by \citet{Tuomi12} in a recent preprint emerged during refereeing of our
paper. In this work, the authors also used an autocorrelated noise model with
exponentially decaying correlation, similarly to our work. Their main conclusion differing
from ours is that using the Bayesian approach they can confirm the existence of the fourth
planet, \emph{d}, with larger statistical evidence. We must admit that this did not
dissolve our criticism concerning this planet. We are of the opinion that such a
difference between our conclusions was induced by different subjective prior distributions
adopted in \citep{Tuomi12}, rather than on the objective information hidden in the RV
data. \citet{Tuomi12} assumed the prior p.d.f. for planetary orbital periods as $\propto
1/P$, meaning a flat distribution in $\ln P$. It is easy to show that the periodogram
approach used in our paper (which basically belongs to the family of AMLET tools)
implicitly assumes a prior which is roughly flat in the \emph{frequency}, implying the
period distribution law $\propto 1/P^2$. Clearly, even if we leave behind the scene the
discussion of what prior is better, the first prior dramatically shifts any data analysis
in favour of longer-period signals, so there is no surprise that \citet{Tuomi12} obtained
more significance for GJ581~\emph{d} than we did. This difference is mostly subjective,
however: we could reach roughly the same effect simply by dealing with renormalized
periodograms, multiplying them by the value of the period in the abscissa.

What concerns the question which prior is better, the answer depends on the purpose of the
analysis and other conditions. If our goal was to analyse a large array of datasets for
many targets and to maximize the outcome of this massive analysis, then we would use the
prior $1/P$, because we know that the period distribution of real exoplanets is much more
uniform in the log-period/log-frequency scale than in the linear-frequency one. However,
here we deal with a planetary system that has high individual importance. For an
individual dataset, the prior $1/P$ may induce some uncomfortable side effects. In
particular, it favours to selection of longer-period aliases instead of the true periods.
A practical example is provided by the two-planet system of HD208487. The Bayesian
analysis assigns a $909$~day period for the second planet of this system
\citep{Gregory07a}, but in the usual periodogram of the residuals there are actually
\emph{two} main peaks at $27$~days and $\sim 1000$~days \citep{Wright07}. These two peaks
can be interpreted as monthly aliases of each other, and offer almost the same best-fit
residuals r.m.s. However, the $1/P$ prior used in \citep{Gregory07a} just suppressed the
shorter-period peak, allowing it to slip away from the view. In the case of GJ581, we must
be even more careful with any pumping of long-period signals, because in such a manner we
can easily pump up some red noise variation that our model was somehow unable to
efficiently eliminate.

Anyway, returning to the planet GJ581~\emph{d}, it is clear that its significance is not
very impressive with the present RV data and, in addition, this significance is highly
model-dependent. Under such circumstances, we prefer to follow the general principle of
the frequentist approach in the statistics, which means that the selection between all
realistic solutions must be done on the worst-case basis. GJ581~\emph{d} needs to be
confirmed by some independent non-HARPS data.

\section*{Acknowledgments}
This work was supported by the Russian Academy of Sciences research programme
``Non-stationary phenomena in the objects of the Universe'' and by the Russian Foundation
for Basic Research, project No. 12-02-31119. I would like to thank Prof. P.~C.~Gregory for
providing an insightful review of this manuscript.

\bibliographystyle{mn2e}
\bibliography{GJ581}

\appendix

\section{Some tips on the maximum-likelihood algorithm}
\label{sec_speed}
\subsection{On the inverse of the noise covariance matrix}
Possibly the fastest way to invert a real symmetric positive-definite matrix like
$\mathbfss V$ is to use the famous Cholesky decomposition: $\mathbfss V = \mathbfss L
\mathbfss L^{\rm T}$, where $\mathbfss L$ is a lower-triangular matrix. It requires
approximately $N^3/6$ floating-point multiplications. Moreover, having the matrix
$\mathbfss L$ at our disposal, we usually do not need to evaluate the inverse $\mathbfss
V^{-1}$ at all, because in true we usually need to evaluate only the matrix-vector
combinations like $\mathbfss L^{-1} \bmath x$ or $\bmath x^{\rm T} \mathbfss L^{-1}$,
which obviously can be obtained using the forward or back substitution. Moreover, these
operations require practically the same CPU time as the direct multiplication by the
precalculated inverse $\mathbfss V^{-1}$ would do.

However, there is a single occurrence where the evaluation via direct matrix inversion
seems the fastest way possible. It is the expression $\trace (\mathbfss V^{-1} \partial
\mathbfss V/\partial p_i)$ in~(\ref{loglik_gradient}). It seems that this task requires
$\sim N^3$ floating-point operations (FLOPs) anyway. However, since this expression must
be evaluated for many parameters $p_i$, it is faster to precalulate $\mathbfss V^{-1}$
based on the Cholesky decomposition (this inversion requires $N^3/3$ floating-point
multiplications) and then to evaluate the necessary matrix trace directly. Notice that the
trace of a matrix product $\trace \mathbfss A \mathbfss B$ is equal just to the scalar
product of the matrices involved, $\sum_{i,j} A_{ij} B_{ij}$, and thus requires only $\sim
N^2$ operations.

\subsection{Avoiding matrix multiplications}
Let us consider the calculation of $F_{p_i p_j}$ in~(\ref{fisher}). Even if we have
precalculated the inverse $\mathbfss V^{-1}$ or use some decomposition of $\mathbfss V$
that makes its inversion easy, the expression~(\ref{fisher}) involves a few matrix
multiplications of very large ($N\times N$) matrices, which require $\mathcal O(N^3)$
FLOPs. This is unsatisfactory and motivates us to find another representation for $F_{p_i
p_j}$ that could be evaluated more quickly. Using the general identity $\bmath x^{\rm T}
\mathbfss A \bmath x = \trace (\mathbfss A \bmath x \bmath x^{\rm T})$ and the relation
$\expect(\bmath r \bmath r^{\rm T}) = \mathbfss V$ (the equality is exact because we
should take the mathematical expectation at the true values of the parameters), we can
transform the first of the expressions~(\ref{loglik_hesse}) as follows:
\begin{eqnarray}
F_{p_i p_j} &=& - \expect \frac{\partial^2 \ln \mathcal L}{\partial p_i \partial p_j} = \nonumber\\
&=&
- \frac{1}{2} \trace \left[ \mathbfss V^{-1} \left(\frac{\partial \mathbfss V}{\partial p_i}
\mathbfss V^{-1} \frac{\partial \mathbfss V}{\partial p_j} -
\frac{\partial^2 \mathbfss V}{\partial p_i \partial p_j} \right) \right] + \nonumber\\
& & + \expect \left[ {\bmath r}^{\mathrm T} \mathbfss V^{-1} \left(
\frac{\partial \mathbfss V}{\partial p_i} \mathbfss V^{-1}
\frac{\partial \mathbfss V}{\partial p_j} -
\frac{1}{2} \frac{\partial^2 \mathbfss V}{\partial p_i \partial p_j}
\right) \mathbfss V^{-1} \bmath r \right] = \nonumber\\
&=& - \frac{1}{2} \trace\left( \mathbfss V^{-1} \frac{\partial \mathbfss V}{\partial p_i}
\mathbfss V^{-1} \frac{\partial \mathbfss V}{\partial p_j} \right) + \nonumber\\
& & + \expect \left( {\bmath r}^{\mathrm T} \mathbfss V^{-1}
\frac{\partial \mathbfss V}{\partial p_i} \mathbfss V^{-1}
\frac{\partial \mathbfss V}{\partial p_j} \mathbfss V^{-1} \bmath r \right)
\end{eqnarray}
Performing the same transform leading to a matrix trace once again, we obtain the final
expression for $F_{p_i p_j}$ in~(\ref{fisher}). Now, what if we apply this last transform
in the opposite direction? Then we obtain the following approximation:
\begin{equation}
F_{p_i p_j} \simeq \frac{1}{2} {\bmath r}^{\mathrm T} \mathbfss V^{-1}
\frac{\partial \mathbfss V}{\partial p_i} \mathbfss V^{-1}
\frac{\partial \mathbfss V}{\partial p_j} \mathbfss V^{-1} \bmath r,
\label{Fp}
\end{equation}
which has a relative error of the order of $1/\sqrt N$ (appearing because we also removed
the expectation operator).

Since we use the Fisher matrix just as a handy approximation of the Hessian with the same
relative error of $\mathcal O(1/\sqrt N)$, the approximation in~(\ref{Fp}) is no worse.
However, (\ref{Fp}) can be evaluated without use of matrix multiplications. Already having
the Cholesky decomposition $\mathbfss V = \mathbfss L \mathbfss L^{\rm T}$, we can easily
perform the first matrix-vector multiplication $\mathbfss V^{-1}
\bmath r = (\mathbfss L^{-1})^{\rm T} \mathbfss L^{-1} \bmath r$ (only $\sim N^2$ FLOPs).
After that we need to perform yet a few matrix-vector multiplications to form
$\partial\mathbfss V/\partial p_i \mathbfss V^{-1} \bmath r$ for all $i$ (also $\sim N^2$
FLOPs). Then we need to multiply these vectors by $(\mathbfss L^{-1})^{\rm T}$ from the
left side (again $\sim N^2$ FLOPs) and evaluate the pairwise scalar products of the
resulting vectors to obtain $F_{p_i p_j}$ for all $i$ and $j$ (only $\sim N$ FLOPs). This
optimized procedure requires only $\mathcal O(N^2)$ FLOPs instead of the original
$\mathcal O(N^3)$ FLOPs.

\subsection{Profit from the matrices sparseness}
Since the noise correlation timescale that we are dealing with is about $10$~days, while
the total time span has the order of $10^3$~days, most elements in the matrix $\mathbfss
V$ are close to zero. Therefore it is highly desirable to set small off-diagonal elements
to zero exactly, and apply some algorithm of Cholesky decomposition and/or inversion tuned
for sparse matrices. It is important that the first thing must be done in a smooth manner:
we cannot just abruptly set all correlations below some small level to zero, since for the
sake of smooth work of our Levenberg-Marquardt algorithm, we need to have continuously
varying gradient and Hessian of the likelihood function. We reach this goal by means of
the following smooth replacement in the argument of the noise correlation function:
$\rho'(x)=\rho(x'(x))$, where $x'=x/(1-(x/x_0)^2)$ and $x_0$ is such that $\rho(x_0)$ is
equal to some small value, e.g. $0.01$. For $x>x_0$ we set $\rho'(x)\equiv 0$. After such
modification, most of the elements in $\mathbfss V$ become exact zeros. Interestingly,
after that we noted a remarkable speed-up of various linear algebra calculations, even
with no use of any specialized sparse-matrix algorithms. This indicates, probably, that
modern CPUs execute various floating-point commands faster when one of the arguments is
zero. With the use of algorithms tuned for sparse matrices, the performance increases even
more dramatically. We unfortunately cannot give any reference or recommendation of any
relevant software package, since the algorithms that we used in this paper we programmed
ourself.

\section{Monte Carlo simulation schemes used in the paper}
\label{sec_simul}
\subsection{Plain Monte Carlo assuming Gaussian noise}
\label{subsec_MC}
\begin{enumerate}
\item First of all, select some mock ``true'' values of the model parameters somewhere in
the region of interest. We may select, for example, the nominal ones given in
Table~\ref{tab_GJ581_bcde_white} for the white-noise model or analogous best fitting
values for the red-noise models, although such choice is not mandatory.

\item Given the chosen vector of ``true'' parameters, evaluate the ``true'' RV values and
the compound RV errors variances (and also correlations for red-noise models).

\item Construct a simulated RV dataset by means of adding to the evaluated RV curve the
simulated Gaussian errors, generated on the basis of previously evaluated uncertainties
and correlations.

\item Based on simulated dataset, evaluate the value of the likelihood function at the
true parameter values from step~1, and the maximum value of the likelihood function for
this trial. Based on these two values, evaluate the modified likelihood ratio statistic
$\tilde Z$ for this trial, which is defined in \citep{Baluev08b}.

\item Save the newly generated value of $\tilde Z$, as well as the set of the simulated
best fitting parameters (when necessary), and return to step~3, if the desired number of
trials has not been accumulated yet.
\end{enumerate}

\subsection{Bootstrap simulation}
\label{subsec_bstrp}
\begin{enumerate}
\item Evaluate the best fitting model and the resulting RV residual.
\item Apply random shuffling procedure separately to the HARPS and Keck sets of the
residuals.
\item Evaluate the statistic $\tilde Z$ and best fitting parameters in the same manner as
in the plain Monte Carlo simulation.
\item Save the resulting value of $\tilde Z$ and parameters and return to step~2.
\end{enumerate}
Note that the boostrap simulation is meaningful only when it is used with a white-noise RV
model, because random shuffling of the residuals basically destroys any correlational
structure of the RV noise, which a red-noise model tries to deal with.

\subsection{Genuinely frequentist Monte Carlo simulation}
\label{subsec_FMC}
\begin{enumerate}
\item Select an $i$th trial point in the space of model parameters $\bxi$ (or residing
inside some given parametric domain).
\item Run the algorithm~\ref{subsec_MC} assuming that true parameters correspond to the
selected point.
\item Save the simulated distribution $P_i(\tilde Z)$ of the test statistic of interest
($\tilde Z$ in our case) and return to step~1.
\item When a sufficiently dense coverage of the mentioned in step~1 parametric domain is
reached, evaluate the function $P(\tilde Z)=\min P_i(\tilde Z)$.
\end{enumerate}
After that, the rigorous frequentist false alarm probability associated with an
\emph{observed} value $\tilde Z_*$ (that was obtained using exactly the same models that
were used during the simulation) can be calculated as $1-P(\tilde Z_*)$. This is a
worst-case assumption method, in other words. Note that if we would stand on the Bayesian
ground, we would evaluate, basically, some weighted average of $P_i(\tilde Z)$ instead of
the minimum, and this would force us to assume some prior distribution of the parameters.
Obviously, in the frequentist approach we need only to circle a parametric domain, since
any prior density inside this domain does not play any role when we find the minimum.

\section{AMLET applicability to the GJ581 case}
\label{sec_AMLET}
Let us first freshen in brief a few practical things that AMLET includes:
\begin{enumerate}
\item The asymptotically unbiased estimations of the model parameters are provided by the
position of the maximum of the likelihood function.

\item Asymptotically, these estimations follow the multivariate Gaussian distribution
with the covariance matrix expressed (again asymptotically) as the inverse of the Fisher
information matrix, defined in~(\ref{fisher_def}).

\item To test some simple ``null'' model against a more complicated alternative one
(which encompasses the null hypothesis as a partial case), we need to construct the
relevant likelihood ratio statistic, and evaluate the false alarm probability associated
with the null hypothesis rejection. The latter false alarm probability can be found from
the known asymptotic $\chi^2$ distribution of the likelihood ratio logarithm.

\item Consequently from the previous point, the multi-dimensional confidence regions for
some set of model parameters are outlined as level curves (or level surfaces) of the
likelihood function, considering it after maximization over the rest of free parameters.
The value of the likelihood ratio corresponding to the global maximum and a given level
curve yields the confidence probability of this level curve (again assuming the asymptotic
$\chi^2$ distribution for logarithm of this ratio).
\end{enumerate}

When the fit model is linear or well linearisable, AMLET is accurate already for
relatively small number of observations. When the model non-linearity increase, the
critical number of observations, after which AMLET becomes applicable, appears
impractically large, so that for practical numbers AMLET offers poor precision. Since in
our case of GJ581 we deal with rather complicated non-linear model of the RV data, we
would like to find out, which AMLET proposition we can be used safely under our concrete
circumstances?

We may notice that the AMLET proposition listed above demonstrate different resistance
with respect to a change of variables (re-parametrization of the original model). For
example, assume we have some model parameter $x$, and we make a replacement $x \longmapsto
y=1/x$, treating the old primary fit parameter $x$ as only a derived one. Even if the
distribution of the estimation of $x$ was exactly Gaussian, the analogous distribution of
$y$ may appear completely non-Gaussian. Therefore, rather formal action like a non-linear
change of variables, which did not really alter the functional structure of our original
model, was able to invalidate some of the AMLET propositions. However, some other
propositions, namely those dealing with only maxima of the likelihood function, remained
intact. Indeed, the maximum value of a function is invariable with respect to any change
of the independent variables (at least if this change is a one-to-one mapping), so the
quantities like the likelihood ratio statistic are invariable with respect to such
re-parametrization. This phenomenon is called sometimes as exogenous and endogenous
non-linearity. The exogenous non-linearity does not belong to the physics of the original
task, and depends on human-controllable things like, for instance, the choice of the
system of free parameters, time reference point, coordinate system, etc. The endogenous
non-linearity represents an immanent property of the task and it cannot be eliminated by
any such trick.

\begin{figure*}
\includegraphics[width=0.556\textwidth]{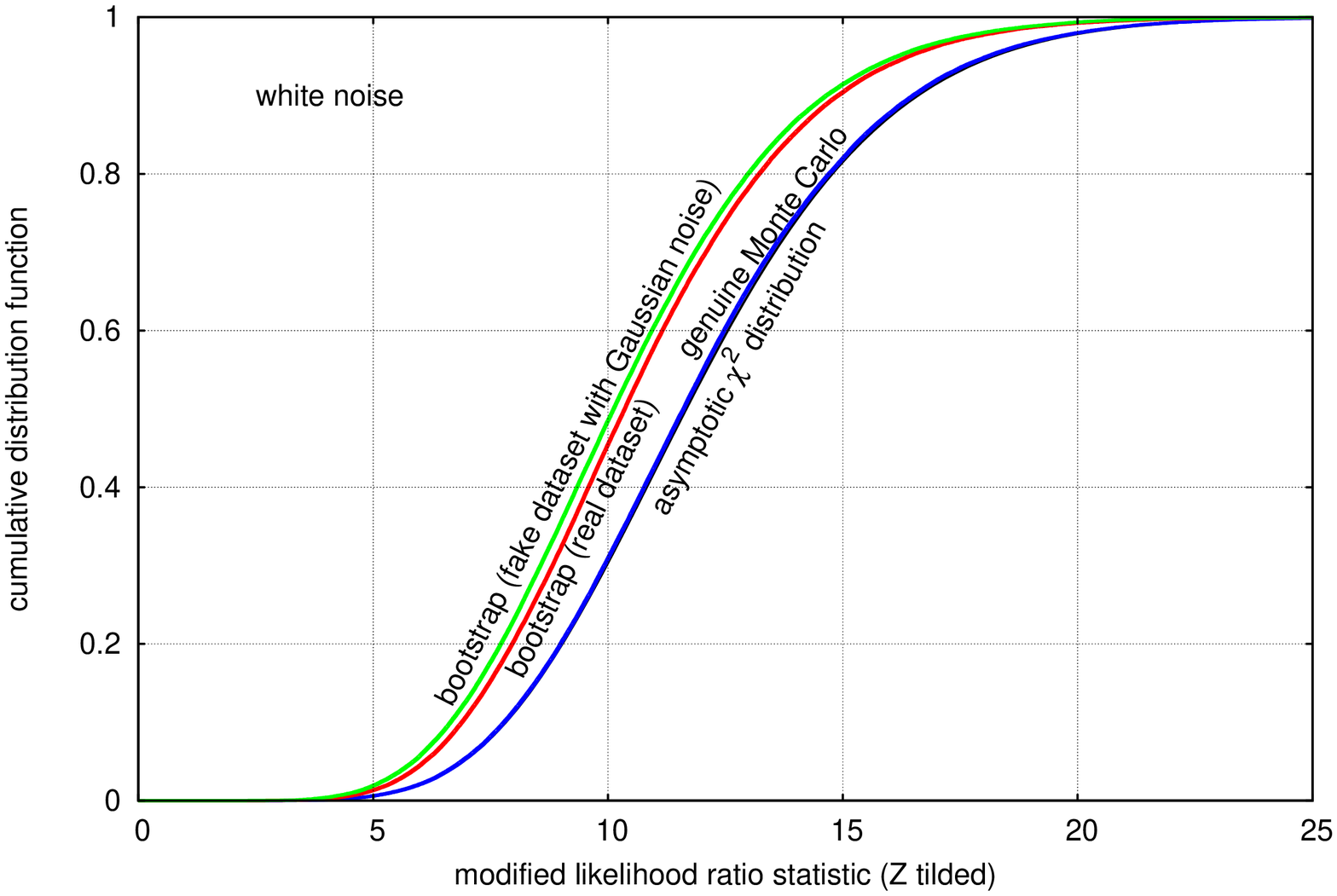}
\includegraphics[width=0.380\textwidth]{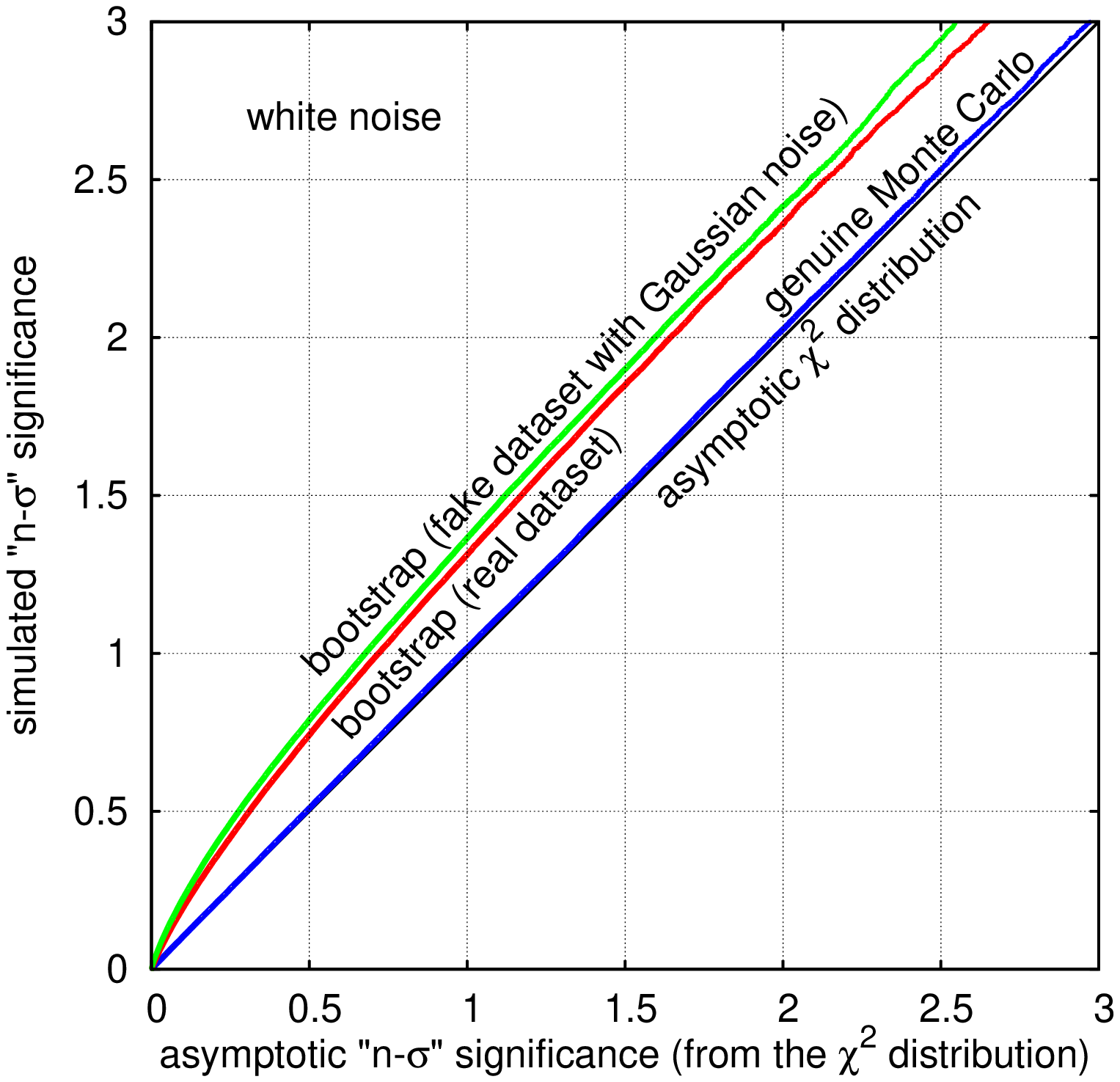}\\
\includegraphics[width=0.556\textwidth]{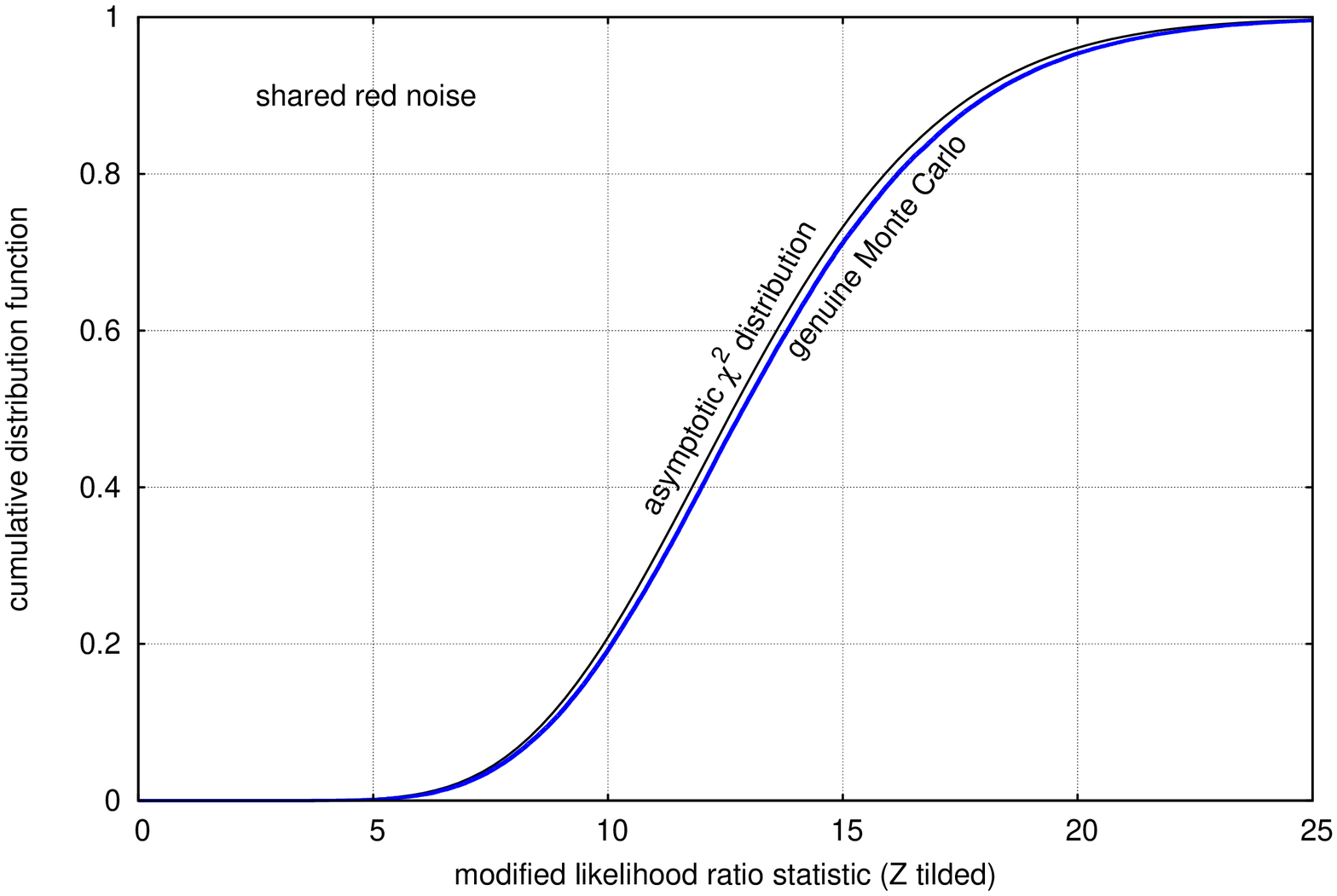}
\includegraphics[width=0.380\textwidth]{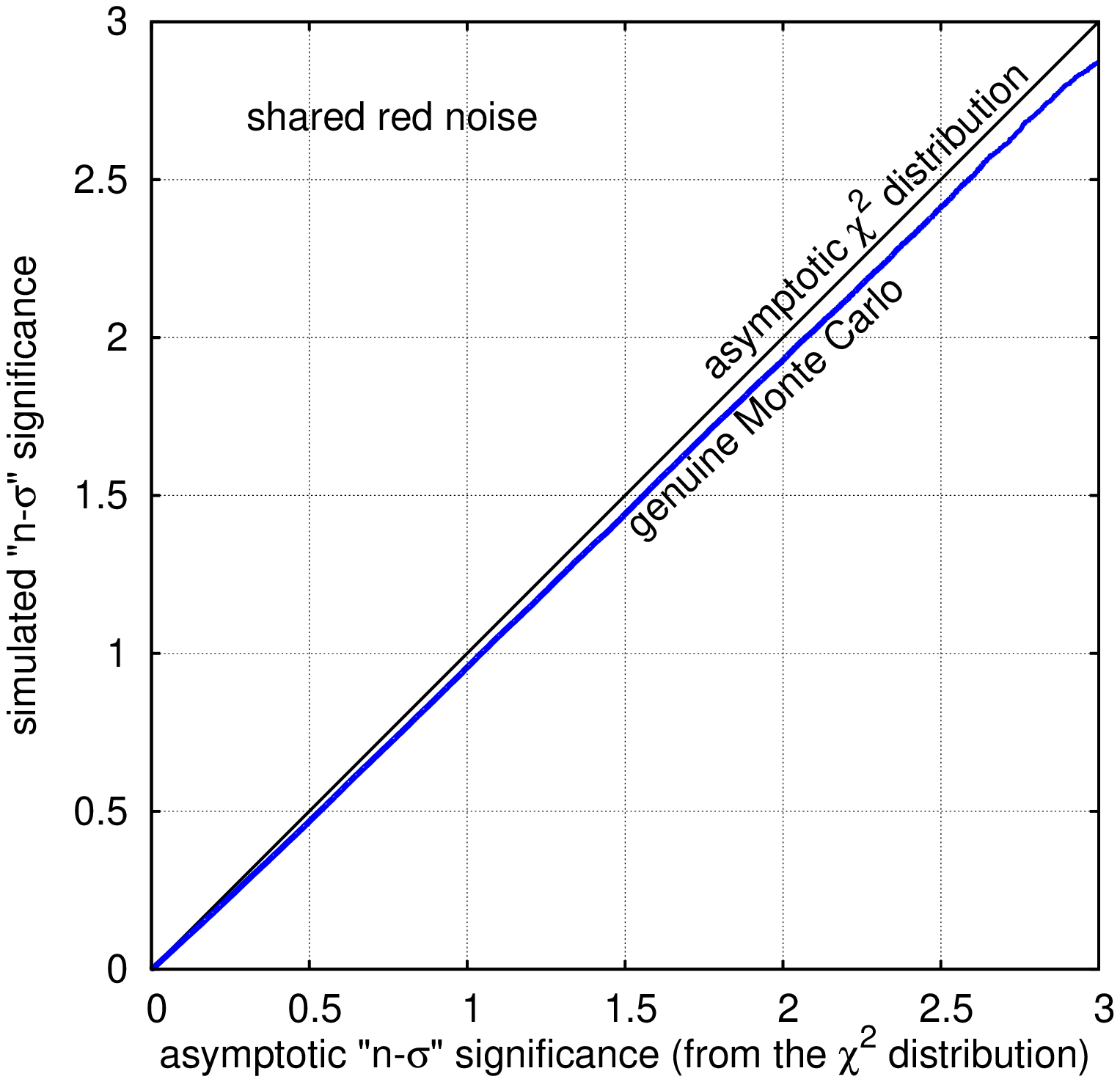}\\
\includegraphics[width=0.556\textwidth]{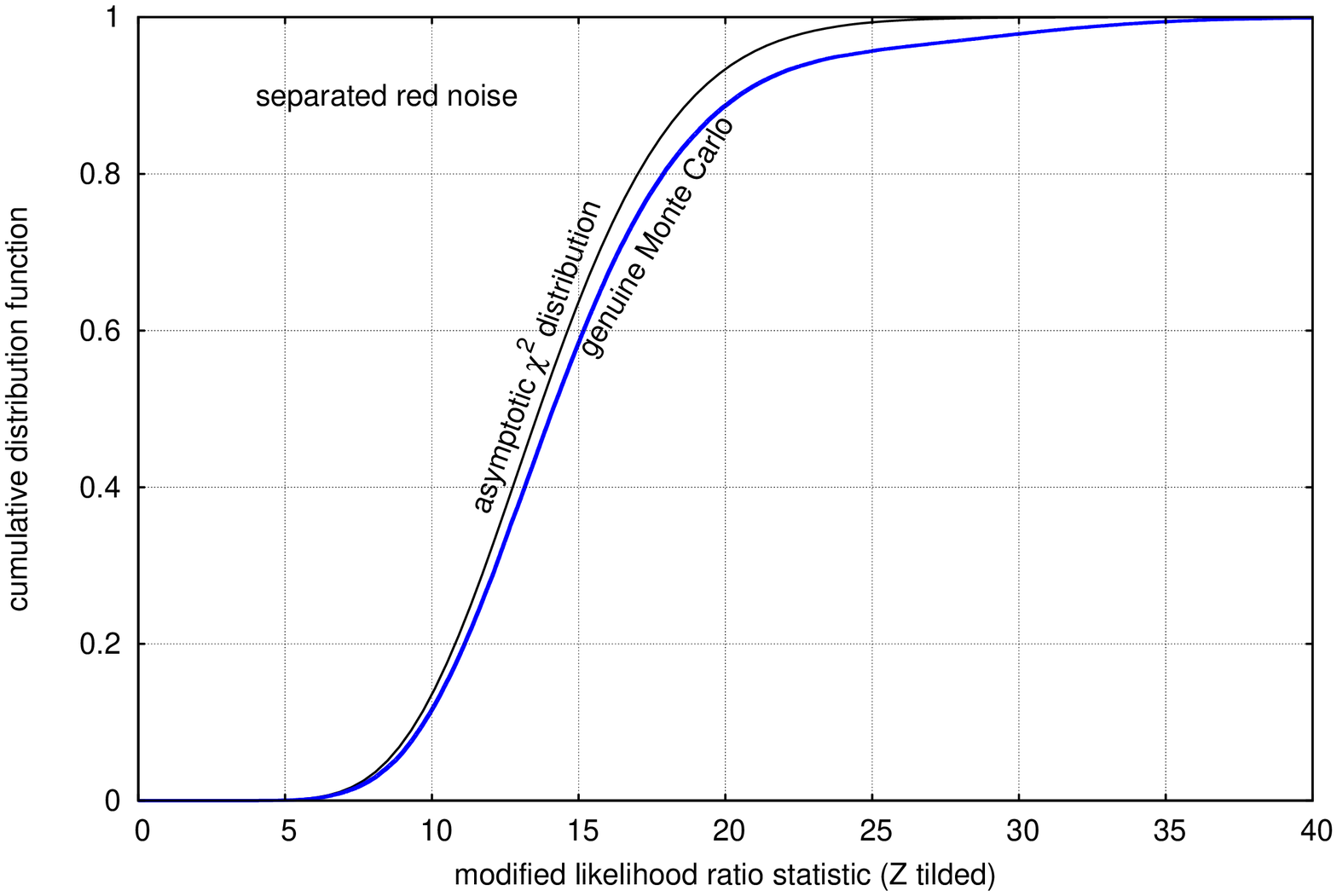}
\includegraphics[width=0.380\textwidth]{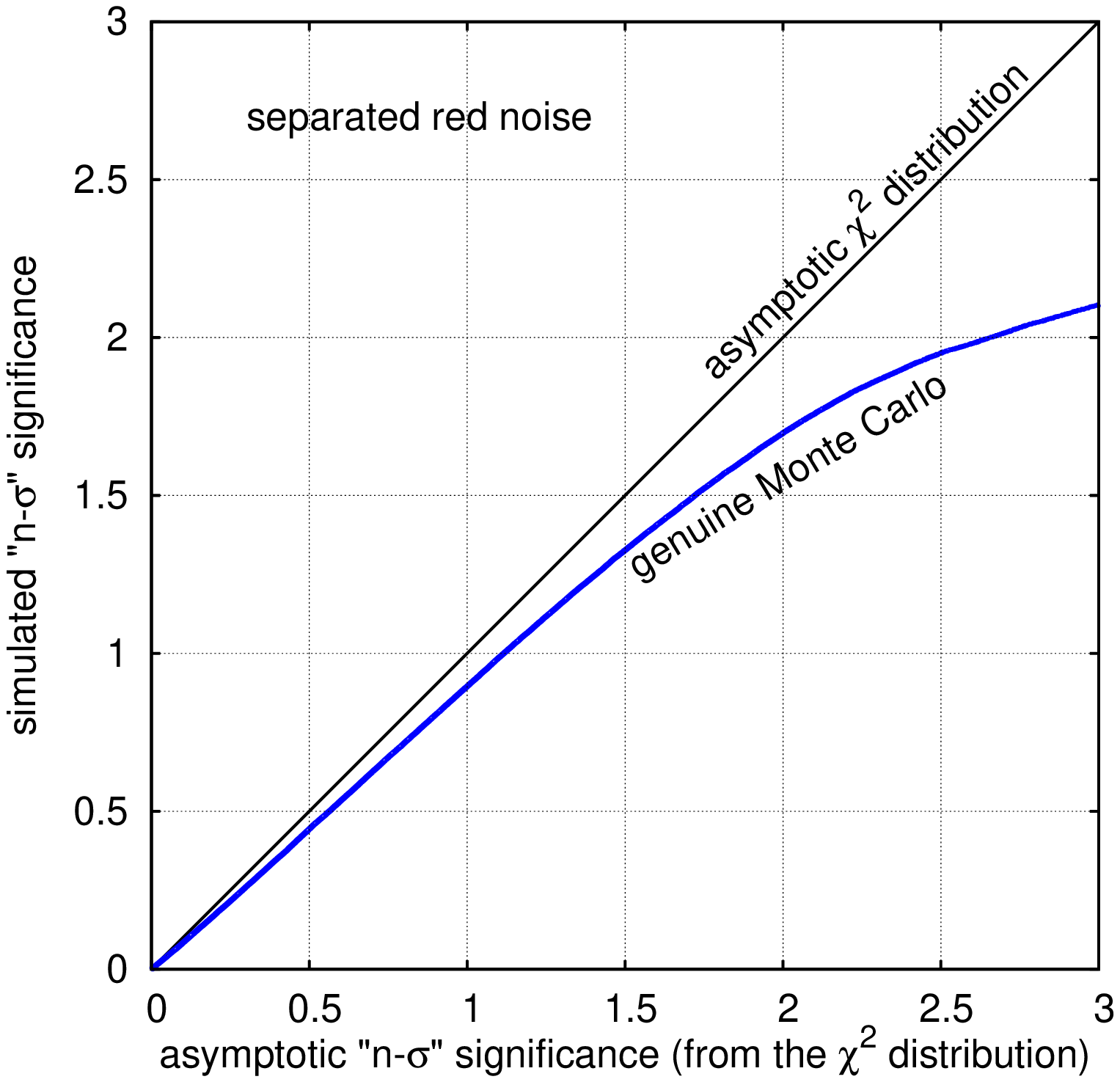}
\caption{Simulated distributions of the statistic $\tilde Z$ from \citep{Baluev08b} and
their asymptotic $\chi^2$ approximation. The null hypothesis was that the model parameters
are equal to the best fitting values (which are treated as true during the relevant
simulation), and the alternative was that all these parameters are unknown (free). The
plots in the left column show the simulated cumulative distributions as functions of
$\tilde Z$, while the relevant simulated significance levels as functions of the
asymptotic $\chi^2$ significance are shown in the right column. The top, middle, and
bottom pairs of panels differ by the assumed RV noise model, according to the marks in
each graph. The RV curve model is always the four-planet one. Note that different noise
models imply different number of degrees of freedom, so the $\chi^2$ distributions become
different too.}
\label{fig_fap}
\end{figure*}

Endogenous part of the non-linearity is the only thing that we really need to take care
of, since anything beyond it is, basically, just a result of incarefully chosen
parametrization. Since in this paper we mainly deal with the likelihood ratio test and its
descendants, we need to check how precisely the asymptotic $\chi^2$ distribution can
approximate the real distribution of the relevant likelihood ratio statistic. We consider
three models to verify: the 4-planet white-noise model, 4-planet shared red-noise model,
and 4-planet separated red-noise model. For each of these models, we perform the Monte
Carlo simulation sequence~\ref{subsec_MC} of the Appendix~\ref{sec_simul}.

For the white-noise model, we also perform two bootstrap simulation series, which is a
popular tool for exoplanetary RV data analysis works \citep[e.g.][]{Marcy05} due to its
resistance to possible non-Gaussian errors in the data. The first simulation is done
according to the scheme~\ref{subsec_bstrp} in the Appendix~\ref{sec_simul}, while in the
second bootstrap simulation we applied the same algorithm to a simulated dataset with
purely Gaussian noise (rather than to the real RV dataset).

The results of these simulations are shown in Fig.~\ref{fig_fap}. For the white-noise
model, we find that the simulated distribution of $\tilde Z$ is in amazingly perfect
agreement with the asymptotic $\chi^2$ approximation. For the shared red-noise model the
agreement is still good. And only for the separated red-noise model AMLET offers poor
precision.

\begin{figure*}
\includegraphics[width=0.75\textwidth]{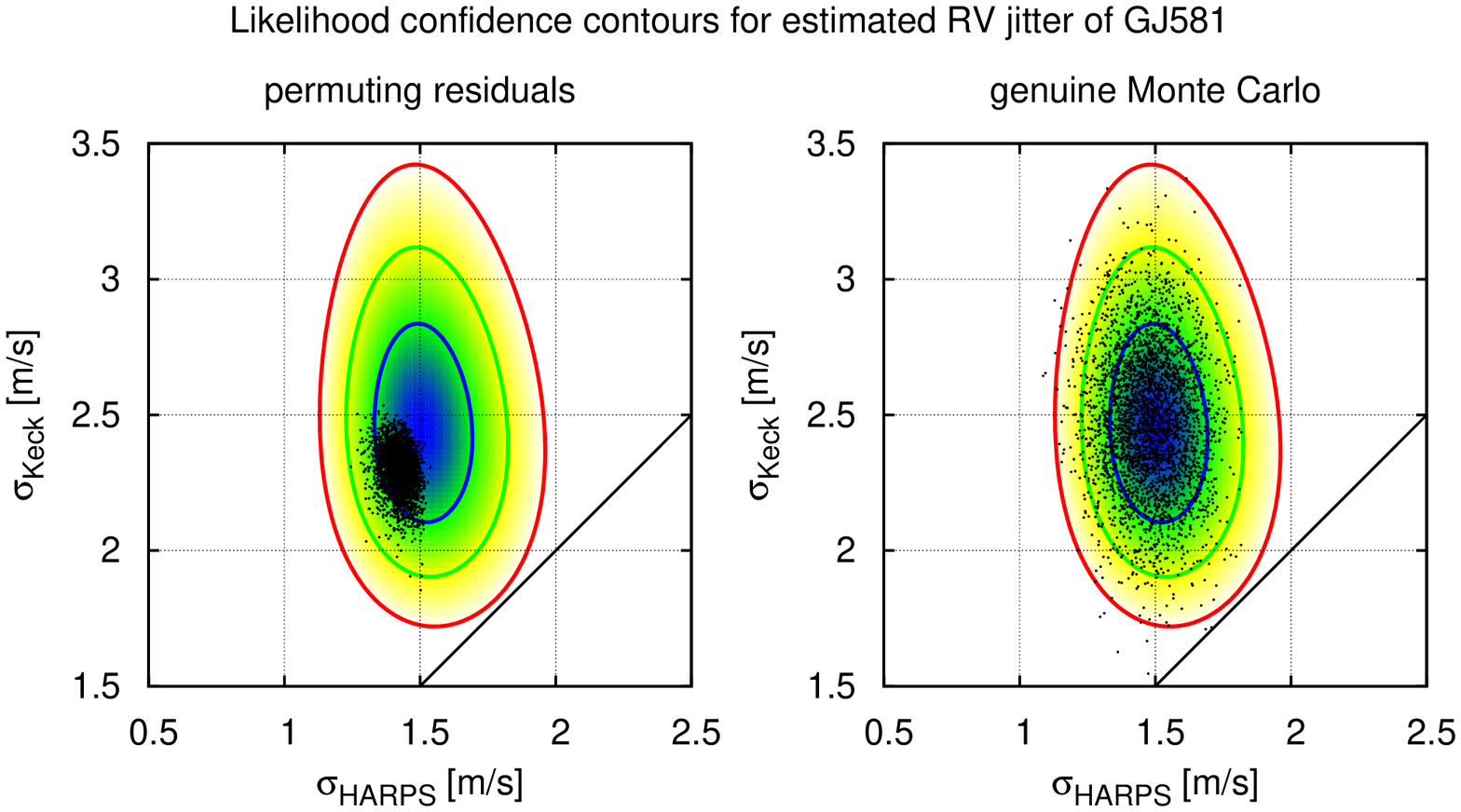}
\caption{Same as Fig.~\ref{fig_lw}, but for the RV jitter parameters of the HARPS and Keck
data.}
\label{fig_wj}
\end{figure*}

It is rather unexpected that in the white-noise case bootstrap simulation disagrees both
with the Monte Carlo and with the $\chi^2$ distribution. This disagreement does not
indicate that the RV noise in the real data is non-Gaussian. Vice versa, two bootstrap
curves lie very close to each other, meaning that real RV measurement errors are
indistinguishable from Gaussian noise. This means that the bootstrap itself is basically
an inadequate tool for our tasks.

So why bootstrap did not work in Fig.~\ref{fig_fap} as we expected? We find that the
reason is contained in the RV jitter parameters. Investigating Fig.~\ref{fig_wj}, we can
see that while the usual Monte Carlo simulations are again in good agreement with AMLET
confidence contours for the HARPS and Keck jitter, the results of the bootstrap are
definitely wrong: they are biased and locked in an inadequately narrow region of the
plane. The reason of this behaviour is clear: while we shuffle the best-fit residuals,
their scatter remains constant, and since the RV jitter is derived mainly from this
scatter, such shuffling keeps both jitter estimations almost constant. This means that
bootstrap cannot be applied to data models involving some parameters of the noise.

\begin{figure*}
\includegraphics[width=0.75\textwidth]{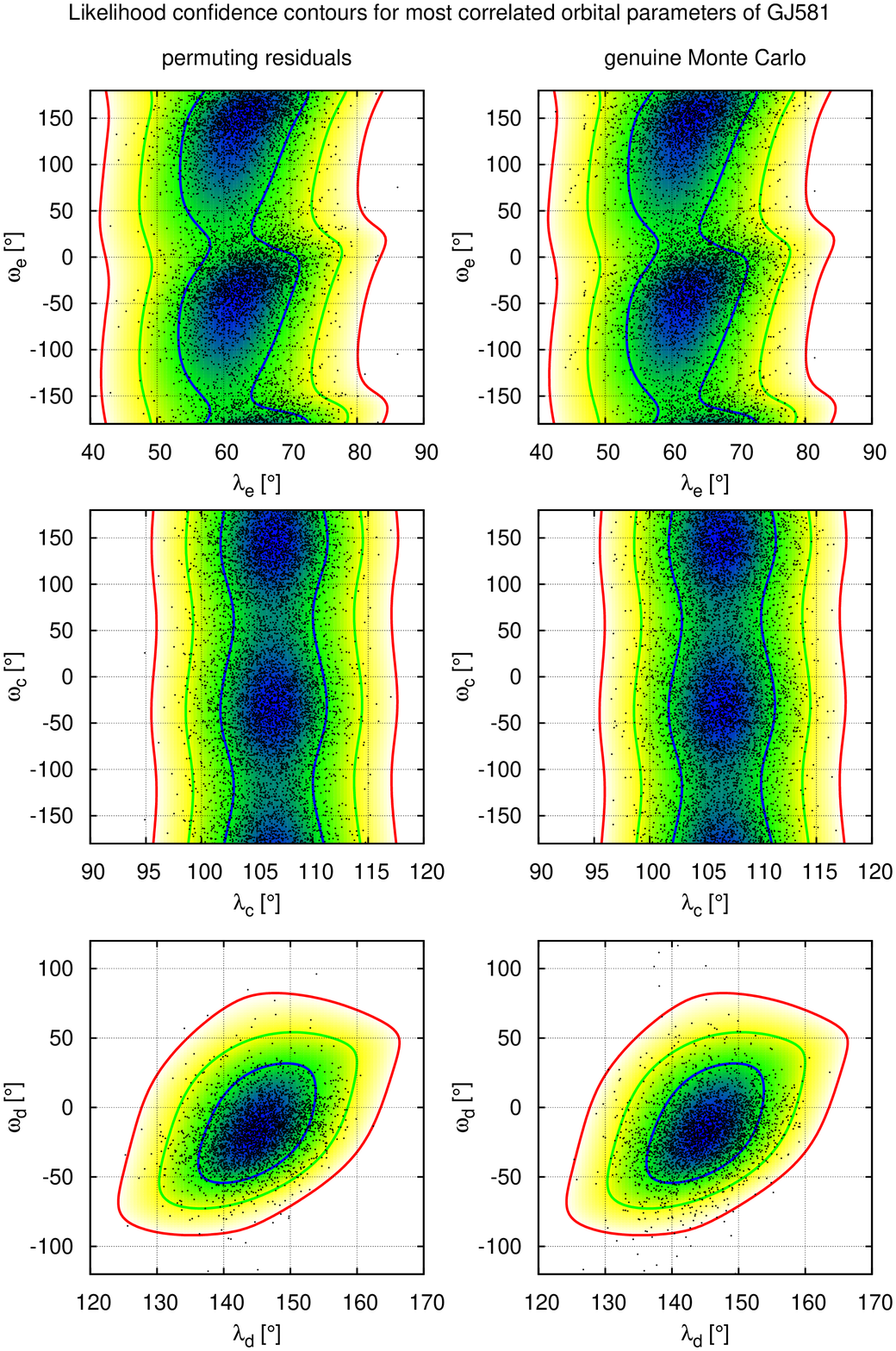}
\caption{The asymptotic confidence contours for a few most correlated pairs of parameters
of the GJ581 4-planet white-noise model, in comparison with Monte Carlo and bootstrap
simulations. These asymptotic confidence contours represent the level curves of the
modified likelihood ratio statistic $\tilde Z$ \citep{Baluev08b}. These contours are shown
by means of colormaps and three reference level curves corresponding to the asymptotic
$1$-, $2$-, and $3$-$\sigma$ significance levels (derived using the asymptotic $\chi^2$
distribution of $\tilde Z$). These contours are identical for the plots in the left and
right columns; the things that differ are the simulated points, that were obtained by
means of the bootstrap (permuting best-fit RV residuals) or pure Monte Carlo (generating
Gaussian noise based on the best-fit model). The number of simulated points shown in each
graph is $3333$.}
\label{fig_lw}
\end{figure*}

Now let us investigate the non-linearity of our RV models in a bit more depth. We consider
2D confidence regions for a few pairs of model parameters. For this goal, we select three
pairs of parameters that demonstrate largest mutual correlations, since such parameters
are usually affected by stronger non-linearity effects \citep{Baluev08c}. As it turned
out, all these pairs involve the mean longitude $\lambda$ and the pericenter argument
$\omega$ of one of the planets. The relevant asymptotic 2D confidence contours,
constructed on the basis of our statistic $\tilde Z$, are shown in Fig.~\ref{fig_lw}.
Clearly, they have little common with ellipses, that we would see for a well-linearisable
model. However, this non-linearity has only exogeneous nature and is caused, obviously, by
small planetary eccentricities, which make the parameter $\omega$ poorly determinable.
Under such circumstance we should better consider, instead of the pairs $(\lambda,\omega)$
the pairs $(\lambda,e\cos\omega)$ and $(\lambda,e\sin\omega)$, since the parameters
$e\cos\omega$ and $e\sin\omega$ are more adequate than $e$ and $\omega$, when dealing with
almost circular orbits.

Our conclusion, that the apparent non-linearity of these parameters is only exogenous, is
confirmed by numerical simulations, which are in good agreement with the asymptotic
confidence contours. The agreement is equally good for the bootstrap and for the pure
Monte Carlo methods, which generate practically identical sets of points. This also
confirms our previous conclusion that the RV data for GJ581 do not show any detectable
non-Gaussianity.

\begin{figure*}
\includegraphics[width=0.556\textwidth]{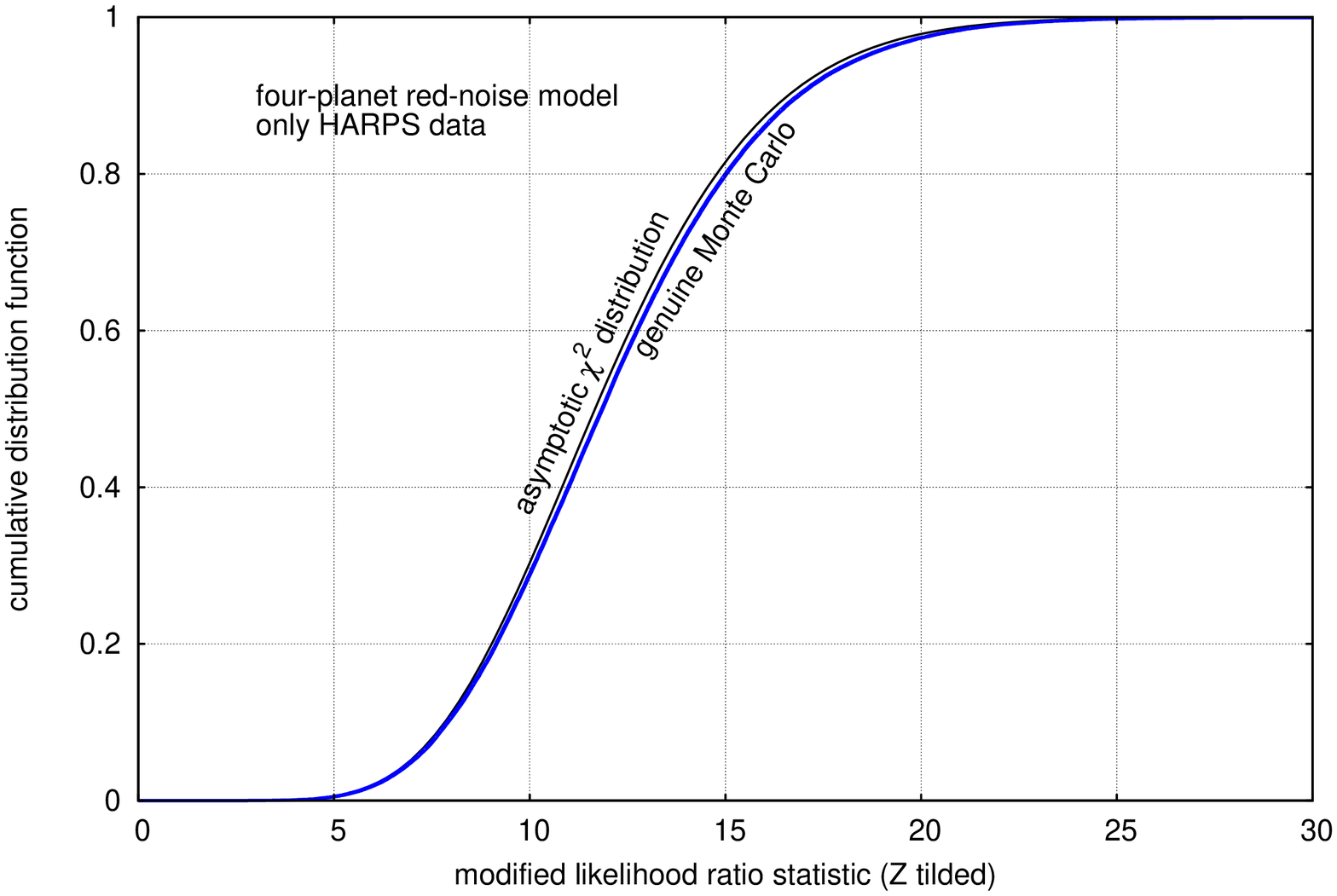}
\includegraphics[width=0.380\textwidth]{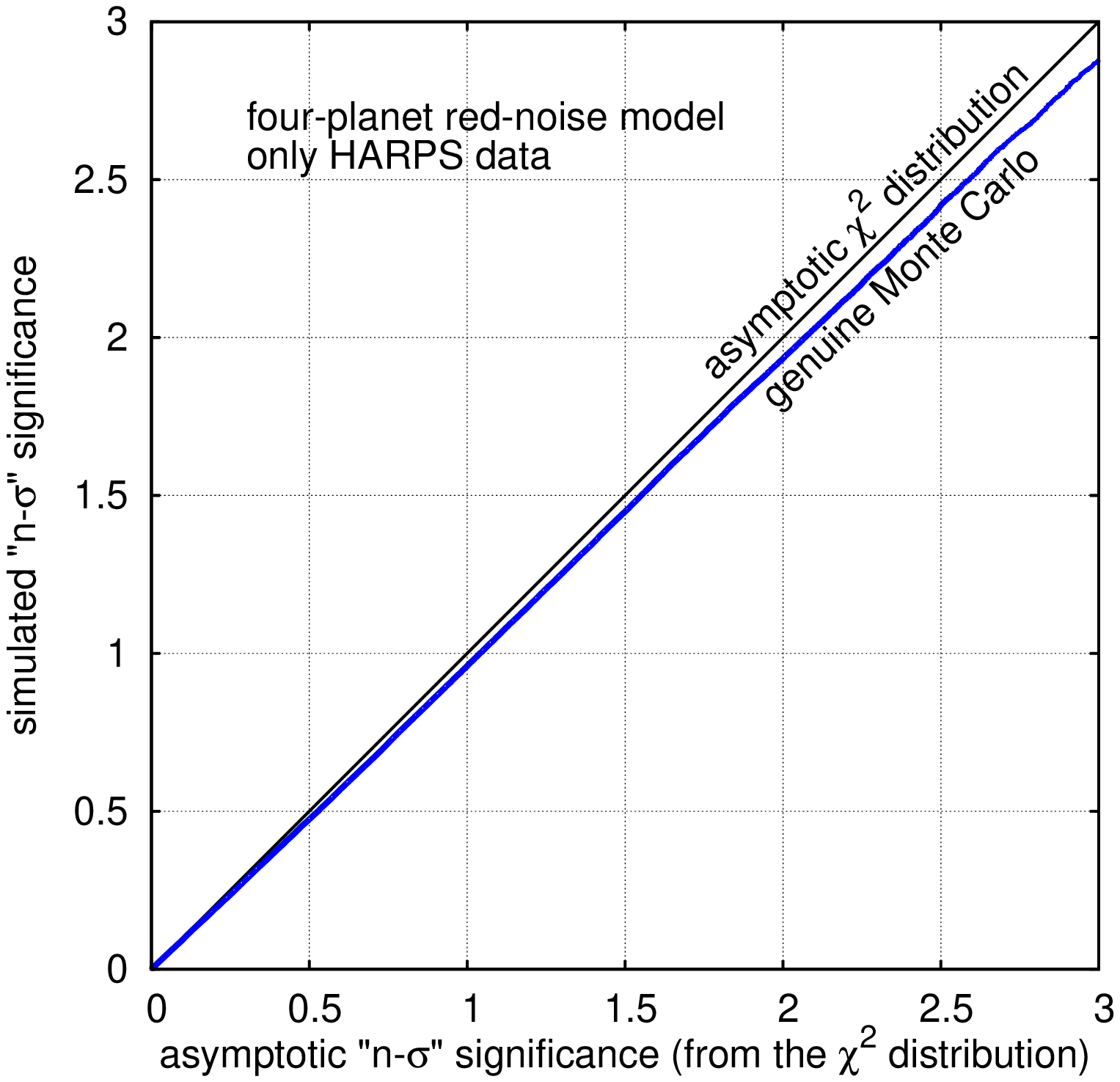}\\
\includegraphics[width=0.556\textwidth]{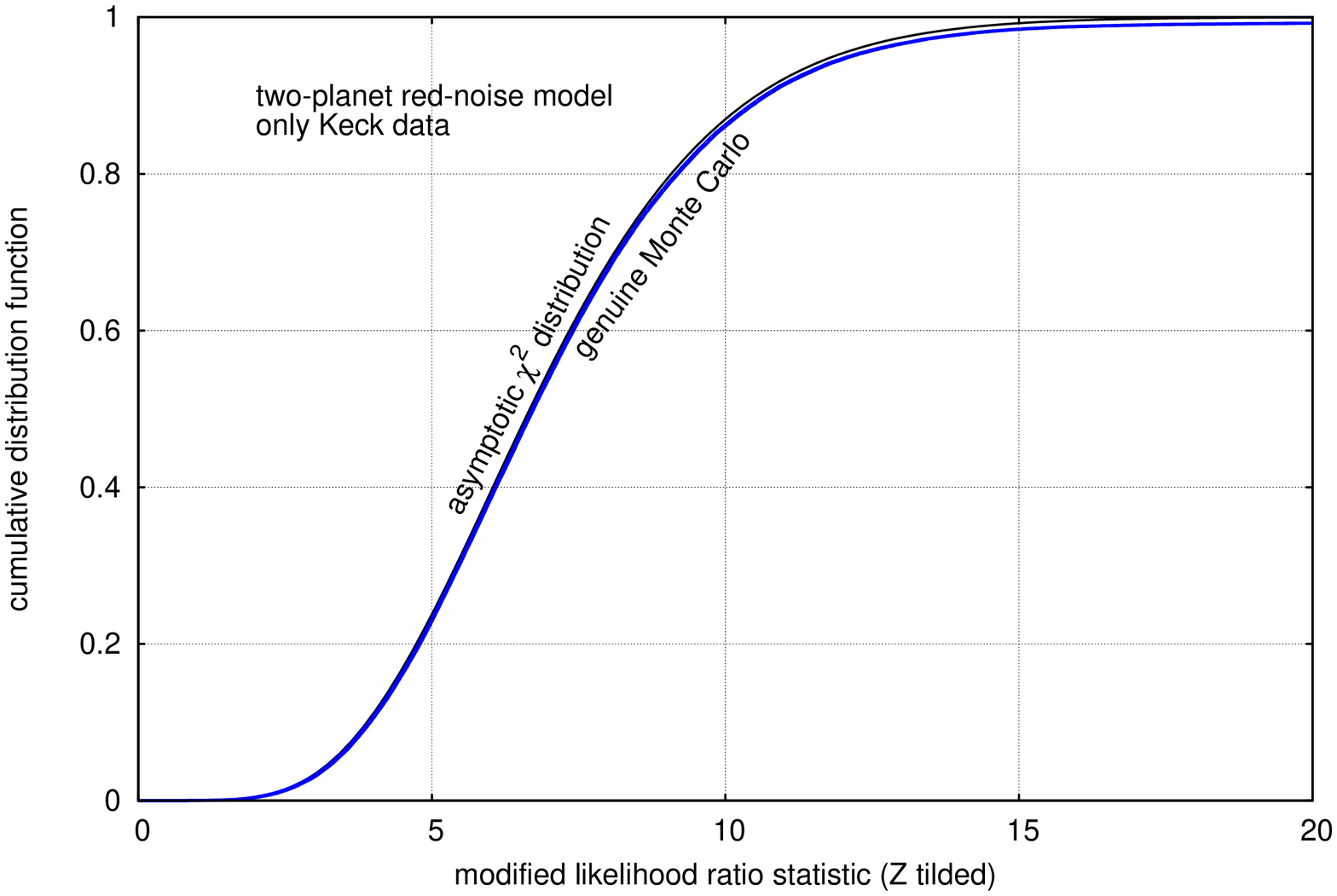}
\includegraphics[width=0.380\textwidth]{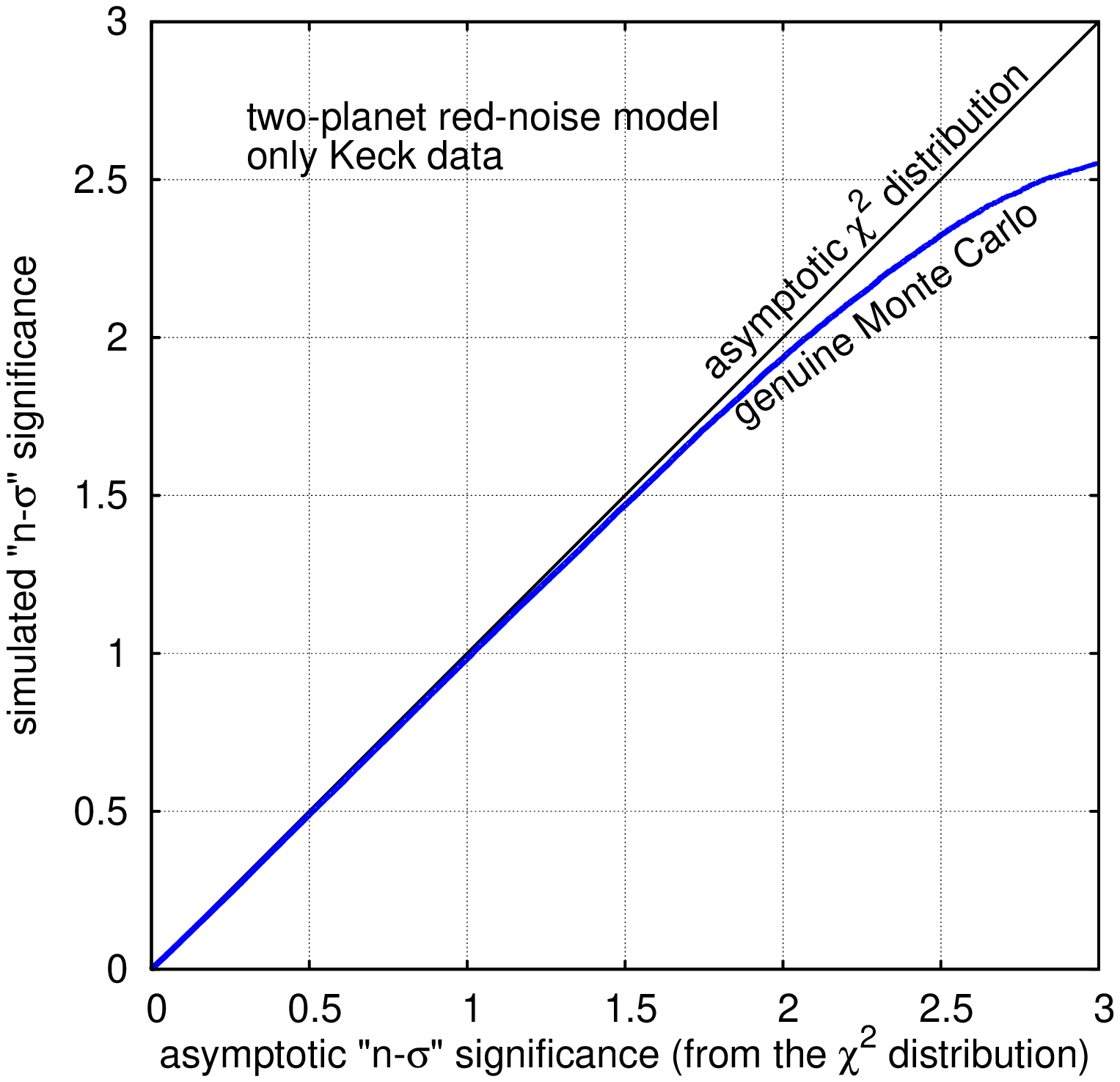}\\
\includegraphics[width=0.556\textwidth]{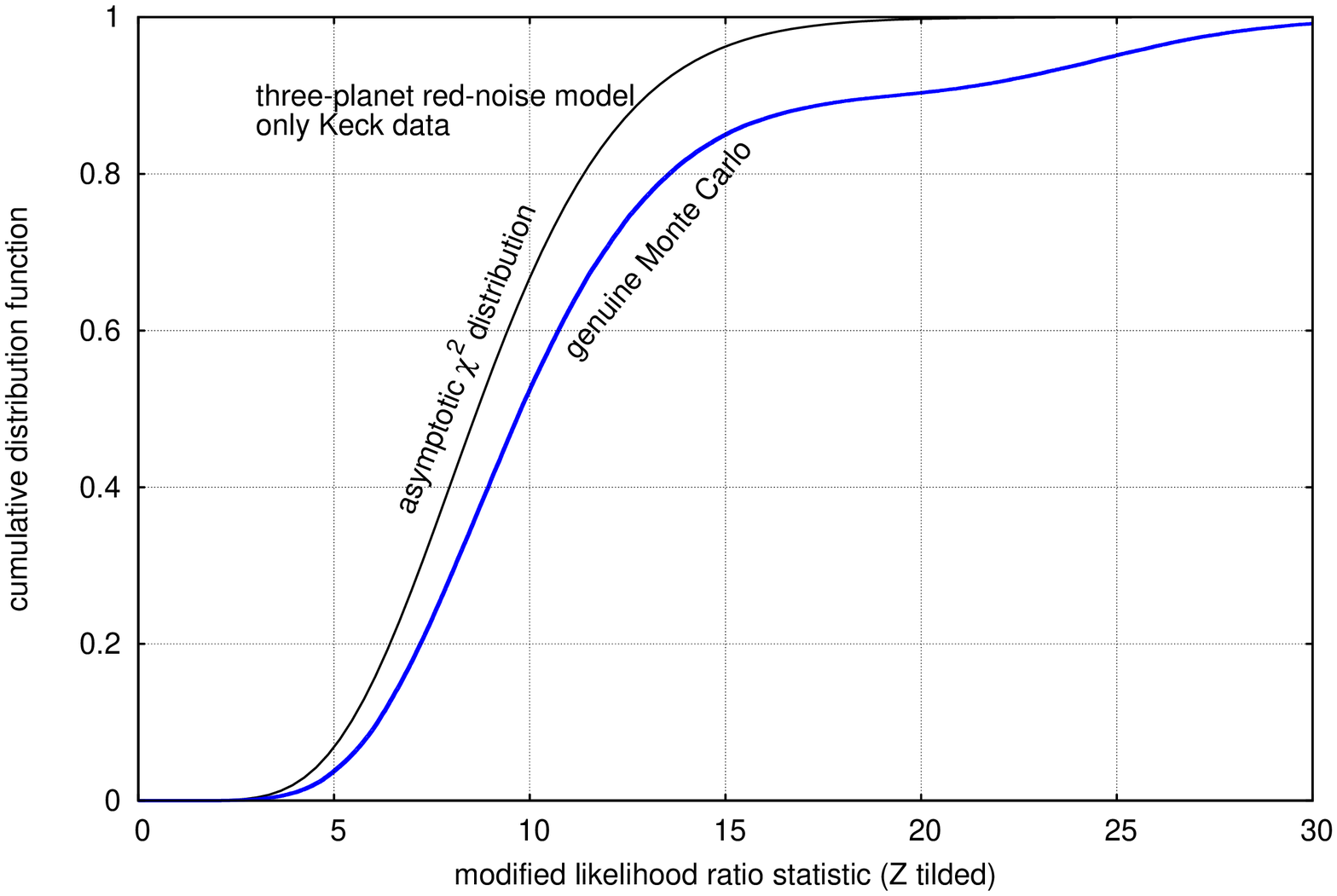}
\includegraphics[width=0.380\textwidth]{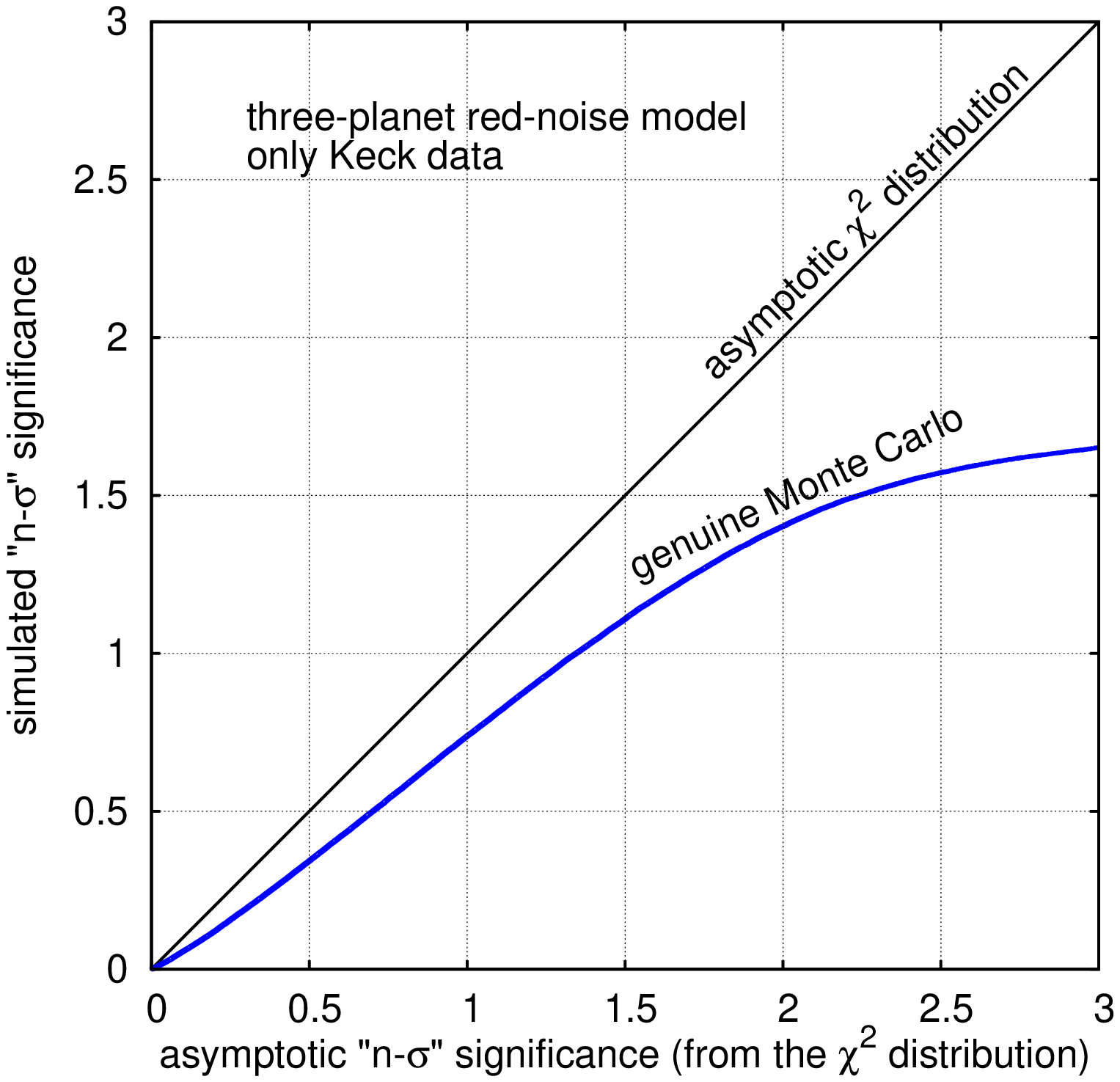}
\caption{Same as Fig.~\ref{fig_fap}, but fitting the HARPS and Keck time series entirely
independently from each other. The top panels correspond to the HARPS dataset with
four-planet model; middle pair~-- to the Keck dataset with two-planet (\emph{b} and
\emph{c}) model; bottom pair~-- to the Keck dataset with three-planet (\emph{b}, \emph{c},
\emph{e}) model. The noise model in each case always incorporated the red component.}
\label{fig_fapsep}
\end{figure*}

And the final question, why the separated red-noise model demonstrates so large deviation
from AMLET's $\chi^2$ distribution in Fig.~\ref{fig_fap}? What is the source of this
endogenous (and thus more importaint) non-linearity? Obviously, the reason is hidden in
the RV noise model, because we have already established that the RV curve model may
produce only negligible endogenous non-linearity. To investigate this question in more
depth, we performed the same Monte Carlo simulation treating the HARPS and Keck data
entirely independently. We again assumed the separated red-noise model for these datasets.
What concerns the RV curve model, for the HARPS data we adopted the four-planet one, while
for the Keck data we considered three-planet and two-planet models (with and without
planet \emph{e}, and with no planet \emph{d}). After that we simulated the distribution of
the statistic $\tilde Z$, and compared it with the relevant asymptotic $\chi^2$
distribution, as in Fig.~\ref{fig_fap}. We obtained that for the HARPS dataset the
agreement is the same good as for the shared red-noise model, while the Keck dataset
demonstrates bad things (Fig.~\ref{fig_fapsep}).

Therefore, the main source making the separated red-noise model statistically poor, is the
Keck dataset, which can provide only rather ill estimations of the RV noise parameters,
when it is used without HARPS data.

\bsp

\label{lastpage}

\end{document}